\documentclass[twocolumn]{aastex631}

\usepackage{amsmath}
\usepackage{subfigure}

\begin{document}

\title{The kinematics of tadpole galaxies at intermediate redshift $z \sim 0.4 - 1.5$}

\author[0009-0008-2970-9845]{Manish Kataria}\thanks{E-mail: manish@iucaa.in}
\author[0000-0002-8768-9298]{Kanak Saha}\thanks{E-mail: kanak@iucaa.in}
\affiliation{Inter-University Centre for Astronomy and Astrophysics, Pune 411007, India.}

\author[0000-0002-1723-6330]{Bruce Elmegreen}
\affiliation{Katonah, NY 10536, USA}
 
\begin{abstract}

Galaxy morphology and kinematics encode complementary information about the assembly history of galaxies, but the extent to which they evolve in tandem remains unclear. Tadpole galaxies, characterized by their distinct head-tail morphology and pronounced asymmetry, provide an ideal laboratory for investigating whether strongly asymmetric galaxies can exhibit ordered galaxy-scale kinematics. In this work, we present a detailed analysis of the ionized-gas kinematics of 20 morphologically selected tadpole galaxies at redshifts $z=0.4 - 1.5$ in the Hubble Ultra Deep Field, using deep VLT/MUSE spectroscopy together with HST and JWST imaging. In 13 out of 20 galaxies, the 2D velocity maps derived from the MUSE [O\,{\sc ii}] $\lambda\lambda3726,3729$ emission show evidence for ordered motion, characterized by moderate velocity gradients along the kinematic major axis, while the remaining systems show no clear evidence for galaxy-scale ordered kinematics. In 16 galaxies, the [O\,{\sc ii}] and rest-frame optical emission, probed by HST/F775W, have spatially coincident centroids, with projected offsets of $<2$~kpc. The morpho-kinematic position angles are also generally well aligned, with a median misalignment of $\Delta{\rm PA}\sim12^{\circ}$. Despite the strongly asymmetric tadpole morphologies, these results indicate that a substantial fraction of these systems retain coherent galaxy-scale gaseous kinematics, suggesting that morphological asymmetry need not imply a globally disordered dynamical state. Together, our analyses suggest that morphological and kinematic settlement does not occur simultaneously and these tadpole galaxies might represent a transient evolutionary phase towards the combined morpho-kinematic settlement process.

\end{abstract}

\keywords{Galaxy evolution (594), Galaxy kinematics (602), Irregular galaxies (790), High-redshift galaxies (734), Galaxy formation (595), Star-forming galaxies (1563)}

\section{Introduction} \label{sec:introduction}

The kinematic state of galaxies provides an important probe of galaxy formation and how they have evolved over cosmic time \citep{Peebles1969, Fall1983, Glazebrook2013, Forster&Wuyts2020}. Spatially resolved observations have shown that coherent galaxy-scale motions are already present at early cosmic epochs \citep{Epinat_etal2012, Contini_etal2016, Guerou_etal2017, Wisnioski_etal2019}. At $z\sim 2$, several surveys have identified velocity gradients and rotation-like kinematics in star-forming galaxies, although these systems generally exhibit larger velocity dispersions and more complex kinematics than galaxies in the nearby Universe \citep{Schreiber_etal2009, Swinbank_etal2009}. Observations with ALMA have extended such studies to even earlier epochs. Using [C{\sc ii}]158 {$\mu$m} emission, the ALPINE survey revealed a diverse range of kinematic states among galaxies at $z = 4 - 6$, including systems consistent with rotation as well as mergers and dispersion-dominated systems \citep{Jones_etal2021, Roman-Oliveira_etal2023}. Higher-resolution observations have subsequently identified regularly rotating systems at $z>6$ \citep{Smit_etal2018, Rowland_etal2024}, while tentative evidence for rotation has also been reported in the $z=14.18$ galaxy, JADES-GS-z14-0, one of the furthest known galaxies \citep{Scholtz_etal2025}. Although ordered galaxy-scale motions appear, ranging from ordered rotation to complex and dispersion-dominated motions, types of kinematic states from ordered rotation to complex and dispersion-dominated motions among high-redshift galaxies \citep{Law_etal2009, Schreiber_etal2009, Wisnioski_etal2015, Wisnioski_etal_2025}. 

This diversity raises a more fundamental question: what determines the kinematic state of a galaxy, and how is it connected to its assembly history and morphology? The presence of a coherent velocity field does not necessarily imply a dynamically cold or rotationally supported system \citep{Shapiro_etal2008}. Galaxies may exhibit ordered motions while retaining substantial turbulent or disordered components, and the relative importance of these components evolves with cosmic time. The emergence and evolution of kinematic order have been quantified through the relative contributions of ordered and disordered gas motions. Most of the star-forming galaxies at $\rm z\sim 1-2$ already show a coherent rotation, but their disks are dynamically hot compared to the local disk galaxies. The $\rm KMOS^{3D}$ survey has found that a large fraction of main-sequence galaxies at $0.7 < z < 2.7$ are rotation-dominated, while their intrinsic ionized-gas velocity dispersions increase strongly with redshift \citep{Wisnioski_etal2015}. Studies following the evolution of galaxies from $z\sim 2$ toward the present day similarly find a progressive transition toward larger $V_{rot}/\sigma$ with rotational support established earlier in more massive systems \citep{Kassin_etal2012, Simons_etal2017}. This evolution is usually described as the epoch of disk settling, in the kinematic sense, during which turbulent, dynamically irregular star-forming systems gradually evolve toward the dynamically colder disks seen in today's Universe. The physical processes governing this evolution, including the acquisition and redistribution of angular momentum, gas accretion and dissipation, and galaxy interactions, mergers, all can affect the kinematic and morphological properties of galaxies in different ways \citep{Danovich_etal2015, Somerville&Dave2015, Pacheco-Ariasetal2026}. Whether kinematic settling is therefore necessarily accompanied by a corresponding morphological transformation remains an open question. Stated in another way, as a galaxy evolves through cosmic time, their morphological and kinematic order need not proceed in tandem. 

The distinction is particularly relevant because the morphology of galaxies also changes strongly with cosmic time \citep{Leeetal2024, Sampaioetal2025}. While the local universe is dominated by relatively symmetric disk and spheroidal morphologies, their high-redshift counterparts, as revealed by the deep HST imaging, are highly asymmetric, irregular, clumpy, and interacting, suggesting a strong departure from dynamical equilibrium \citep{ElmegreenDM_etal2005, Bournaud_etal2007, Ceverino_etal2010, Guo_etal2011, Conselice2014, Kartaltepe_2015, Shibuya_etal2016}. Tadpole and clump-chain galaxies represent an extreme example of such morphological asymmetry, typically characterized by a compact, often intensely star-forming head connected with a comet-like diffuse, elongated tail and multiple star-forming clumps. \cite{Straughn_etal2006} systematically identified these systems in the HUDF and suggested that their disturbed appearance could trace dynamically unrelaxed, early-stage mergers. Subsequent studies, however, indicated that tadpole morphology is not uniquely associated with merging. Some objects may instead be lopsided or clumpy disks dominated by an off-center star-forming region. At the same time, asymmetric gas accretion or interaction with the surrounding medium may provide alternative origins \citep{ElmegreenBG_Elmegreen_DM_etal2010, Straughn_etal2015}. Nevertheless, their greater prevalence at earlier cosmic times makes them a useful galaxy population for investigating the relationship between morphological evolution and kinematic settling. In particular, tadpoles or clump-chain galaxies provide an opportunity to ask whether strongly asymmetric star-forming structures necessarily correspond to disturbed galaxy-scale kinematics, or whether ordered motions can persist despite pronounced morphological asymmetry.

Spatially resolved spectroscopy of nearby tadpole galaxies provides some evidence that strongly asymmetric morphologies can coexist with organized kinematics. \cite{Almeida_etal2013} found signatures consistent with rotation in five of seven local tadpoles, with the inferred kinematic centers often displaced from the bright star-forming heads. The tadpole-heads also frequently showed lower metallicities than the surrounding galaxy, which they interpreted as evidence for recent accretion of metal-poor gas. Together with the multiwavelength HUDF study of \cite{Straughn_etal2015}, which found tadpoles to be generally young, low-mass systems undergoing active assembly, these results show that strongly asymmetric or disturbed morphology can coexist with an organized underlying velocity field \citep{Puech2010}. However, the interpretation of tadpole kinematics is not straightforward because an ordered velocity gradient does not necessarily imply rotation. A velocity gradient across the main body of the galaxy could arise from rotation, but coherent motion along the head-tail structure could also reflect streaming motion, clump movement, gas inflow or outflow, ram-pressure effects, or other non-circular motions. Moreover, the ionized gas emission need not be centered on the stellar light distribution. Consequently, the kinematic center, ionized-gas distribution, and photometric center may differ, and these offsets themselves may contain information about the dynamical state of these unrelaxed systems. 

In this work, we investigate morphologically selected HUDF tadpole galaxies at an intermediate redshift range ($z \sim 0.4 - 1.5$) using deep integral field spectroscopy by MUSE on VLT together with HST and JWST imaging. These tadpole galaxies provide a unique opportunity to directly test whether the elongated morphology traces their gaseous kinematic axis or whether the stellar morphology and ionized-gas motions become significantly decoupled. Starting from the \cite{Straughn_etal2006} catalog, we select systems with detectable [O{\sc ii}] $\lambda\lambda3726,3729$ emission and use higher-S/N pseudo-slit spectra to measure the edge-to-edge velocity shear along the morphological major and perpendicular directions. We characterize their velocity fields and examine the relationship between the spatial distribution of ionized gas, the photometric morphology, and the inferred kinematic structure. We test whether the observed velocity fields are consistent with ordered rotational configurations and identify coherent departures from such models. Due to insufficient spatial resolution, we explicitly account for the effects of the instrumental PSF and LSF when interpreting the observed velocity fields and addressing the broader question of whether these tadpoles are kinematically settled, as well as the possible connection between their asymmetric morphology and kinematic state. 

The rest of the paper is organized as follows: Section~\ref{subsec:data} describes observations and the available archival data. The sample selection criteria and the stellar mass - SFR properties are defined in Section~\ref{subsec:sample}. In Section~\ref{sec:cube_contisub_nb} we show in detail our methods to make continuum subtraction and creation of narrow band [O{\sc ii}] images. We then measure the offset in the HST and [O{\sc ii}] emission in Section~\ref{sec:optical_o2_offset}. To mitigate the low SNR spectra for kinematic fitting, we adopted a pseudo slit method, as can be seen in Section~\ref{sec:Pseudo_slit_extraction}. We also included the 2D kinematics from MUSE data in Section~\ref{subsec:2d_kin_maps} along with the shear measurements from the pseudo slit in Section~\ref{subsec:delv}. The determination of kinematic PA using the rotating pseudo-slits is described in Section~\ref{sec:kin_pa}. Section~\ref{sec:pa_alignment} and \ref{sec:morph_kin_misalignment} describes the $\rm \Delta PA$ definition and misalignment results respectively. Finally, in Section~\ref{sec:conclusion} we state the discussion and conclude our analysis.

Throughout this work, we adopt a flat $\Lambda$ Cold Dark Matter $\rm (\Lambda CDM)$ cosmology \citep{White&Rees1978, Blumenthal_etal1984}, with $\rm H_{0} = 70~km~s^{-1}~Mpc^{-1},~\Omega_{M} = 0.3,~and~\Omega_{\Lambda} = 0.7$

\section{data and source selection} \label{sec:data_sample}

\subsection{Data} \label{subsec:data}

\begin{figure}
    \centering
    \includegraphics[width=\linewidth]{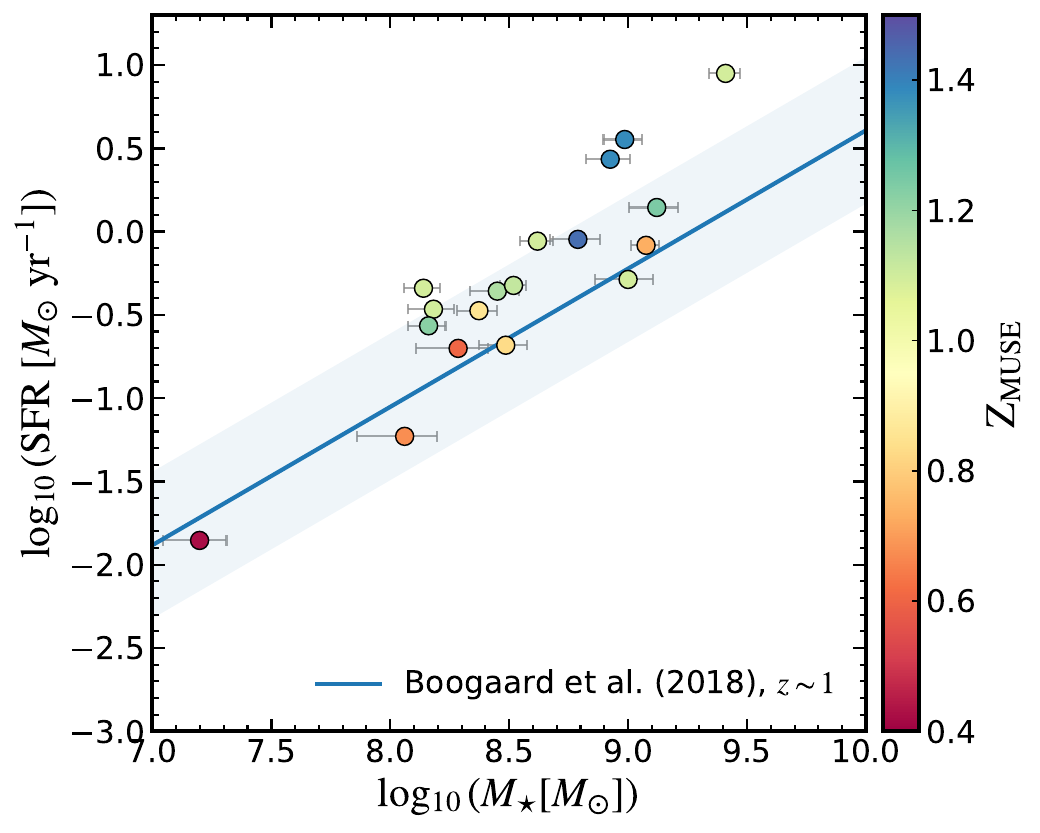}
    \caption{Log star-formation rate as a function of log stellar mass for the tadpole galaxy sample. Points are color-coded by the MUSE spectroscopic redshift, $z_{\rm MUSE}$, and horizontal error bars indicate the uncertainties on the stellar masses. The solid blue line shows the star-forming main-sequence relation from \cite{Boogaard_etal2018} at $z = 1$, with the shaded region indicating its scatter.}
    \label{fig:tpg_main_seq}
\end{figure}

We primarily make use of the MUSE UDF data from the MUSE Deep Survey \citep{Bacon_etal2017, Bacon_etal2023} of the HST GOODS-South Hubble Ultra Deep Field (HUDF; \cite{Illingworth_etal2013}). The HUDF has been observed with MUSE at three different depths, providing a unique combination of area and sensitivity. These include the HUDF Mosaic, with an exposure time of $\sim 10$ hrs; UDF-10, with an exposure time of $\sim 31$ hrs; and the MUSE eXtremely Deep Field (MXDF), with an exposure time of $\sim 140$ hrs. The three datasets cover progressively smaller but deeper regions, with fields of view of approximately $\sim 9$, $\sim 1$, and $\sim 0.8~{\rm arcmin}^{2}$ respectively. MUSE is an optical integral-field spectrograph \citep{Bacon_etal2010, Bacon_etal2014} mounted on Unit Telescope 4 (UT4/Yepun) of the Very Large Telescope (VLT) at the ESO Paranal Observatory. In its wide-field mode, MUSE provides a field of view of ($1^{\prime}\times1^{\prime}$), sampled at $0.2^{\prime\prime}$ per pixel. The instrument covers the optical wavelength range of ($\sim$ 4750 -- 9350~\text{\AA}) in the observed frame, with a spectral sampling of 1.25~\text{\AA} pix$^{-1}$ along the dispersion axis. This combination of extensive spatial coverage, optical wavelength range, and integral-field capability makes MUSE particularly powerful for deep-field studies.

We make use of publicly available high-angular-resolution imaging from the Hubble Space Telescope (HST) and the James Webb Space Telescope (JWST). The HST imaging was obtained from the Hubble Legacy Archive (HLA), while the JWST imaging comes from the JWST Advanced Deep Extragalactic Survey (JADES; PIs: Daniel Eisenstein and Nora Luetzgendorf; \citealt{Eisenstein_etal2026}). These datasets provide deep optical and near-infrared continuum imaging for galaxies in the redshift range ($0.4<z<1.5$) studied in this work. The high spatial resolution of these images is essential for identifying the asymmetric head-tail morphology of the tadpole galaxies, visually inspecting their structures, and defining the morphological axes used for the pseudo-slit kinematic analysis.

\begin{figure*}[!ht]
    \centering
    \includegraphics[width=\linewidth]{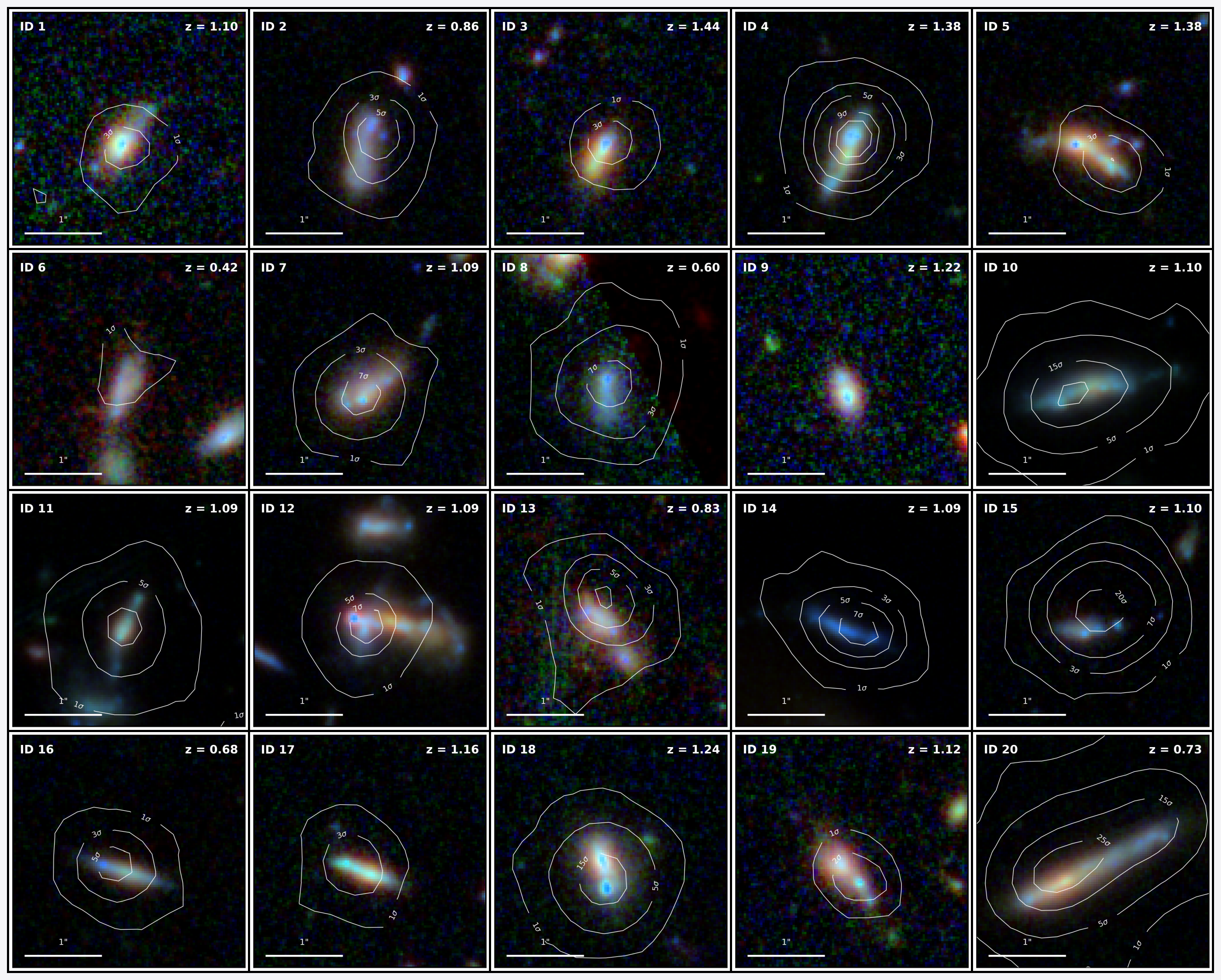}
    \caption{JWST/NIRCAM RGB using F356W(R) + F150W(G) + F090W(B)   composite images of the 20 morphologically selected tadpole galaxies in our sample. Each panel is labeled with the galaxy ID and MUSE spectroscopic redshift. White contours show the continuum-subtracted MUSE [O{\sc ii}] $\lambda\lambda3726,3729$ narrow-band emission, with the contour labels indicating the corresponding significance levels.}
    \label{fig:tpg_kin_rgb}
\end{figure*}

\subsection{Sample selection} \label{subsec:sample}

We take the parent sample from the morphologically selected tadpole galaxy catalog of \citet{Straughn_etal2006}. This catalog was constructed for galaxies in the HUDF using an automated selection algorithm designed to identify the characteristic head-tail morphology of tadpole systems. The algorithm was implemented using SExtractor \citep{Bertin&Arnouts1996} with different parameter configurations to account for the asymmetric light distributions of these galaxies. A compact bright head and an extended lower-surface-brightness tail or clump-chain morphology characterize the selected sources. Such systems are particularly useful for studying galaxy assembly, as their disturbed and highly asymmetric morphologies may be associated with clumpy star formation, tidal interactions, or early disk-building phases.

In the present work, we first cross-match the positions of all galaxies in the Straughn et al. catalog with those in the MUSE UDF catalog from the MUSE UDF survey within a search radius of $0.5^{\prime\prime}$. We first search in the redshift range $z\sim0.27-1.5$, over which the [O{\sc ii}] $(\rm \lambda\lambda\ 3726,3729\ \mathring{\text{A}})$ doublet falls within the MUSE wavelength coverage. The redshift range probes an important epoch, spanning the later stages of cosmic noon and the subsequent transition toward the dynamically colder galaxy population observed in the nearby Universe. The combination of our redshift selection and MUSE spectral coverage yields 37 galaxies out of the original 165 tadpole candidates. We then visually inspect these galaxies using the available HST and JWST imaging to assess their morphology in greater detail. During this step, we remove sources that appear to be obvious regular disks, as well as highly irregular systems where a clear tadpole-like head-tail or clump-chain structure cannot be reliably defined. After this visual inspection, we obtain a sample of 20 galaxies for the kinematic analysis. Some of the galaxies were further dropped in some analysis after inspecting the MUSE spectra around the expected [O{\sc ii}] emission. In a few sources, the [O{\sc ii}] doublet was either undetected or too weak to yield reliable velocity measurements from the pseudo-slit spectra. We therefore exclude these galaxies from the kinematic analysis and include only those sources for which the [O{\sc ii}] emission is detected with sufficient signal-to-noise across multiple spatial apertures.

The stellar mass v/s star-formation rate of the sample is shown in Figure~\ref{fig:tpg_main_seq}, with values taken from the \cite{Santini_etal2015} catalog. We use the 'M\_med' from the catalog for stellar masses, as it combines seven estimates that share a Chabrier IMF \citep{Chabrier2003} and BC03 stellar templates \citep{Bruzual&Charlot_2003}. \cite{Santini_etal2015} conclude that this combined median is generally more reliable than any individual mass estimate. The SFR is taken from the Method 'SFR\_14a' because it uses the same IMF and stellar population model and returns the best fit among the allowed star-formation histories. The galaxies span a relatively low-mass regime, with stellar masses of approximately $\log_{10}(M_\star/M_\odot)\simeq7.5$--$9.5$, although most of the sample is concentrated between $\log_{10}(M_\star/M_\odot)\sim8$ and $9$. Their star-formation rates extend over nearly three orders of magnitude, from $\log_{10}({\rm SFR}/M_\odot\,{\rm yr}^{-1})\sim-1.9$ to $\sim1$. For comparison, we show the star-forming main-sequence relation of \cite{Boogaard_etal2018} at $z=1$, since it was fitted only for $0.11 < z < 0.91$, and for constraining the low mass end of the main sequence with the fitted stellar mass range of $\rm 7 < log(M_{*}/M_{\odot}) < 10$. The galaxies broadly occupy the region expected for low-mass star-forming galaxies over this redshift interval, with many systems lying above the $z=1$ relation, consistent with the increase in typical star-formation activity toward higher redshift. The sample, therefore, primarily probes actively star-forming, low-mass galaxies during the intermediate-redshift phase of galaxy assembly considered in this work.

\section{Continuum subtracted cube and construction of [O{\sc ii}] $(\rm \lambda\lambda\ 3726,3729\ \mathring{\text{A}})$ narrow band images} \label{sec:cube_contisub_nb}

\begin{figure}
    \centering
    \includegraphics[width=0.5\textwidth]{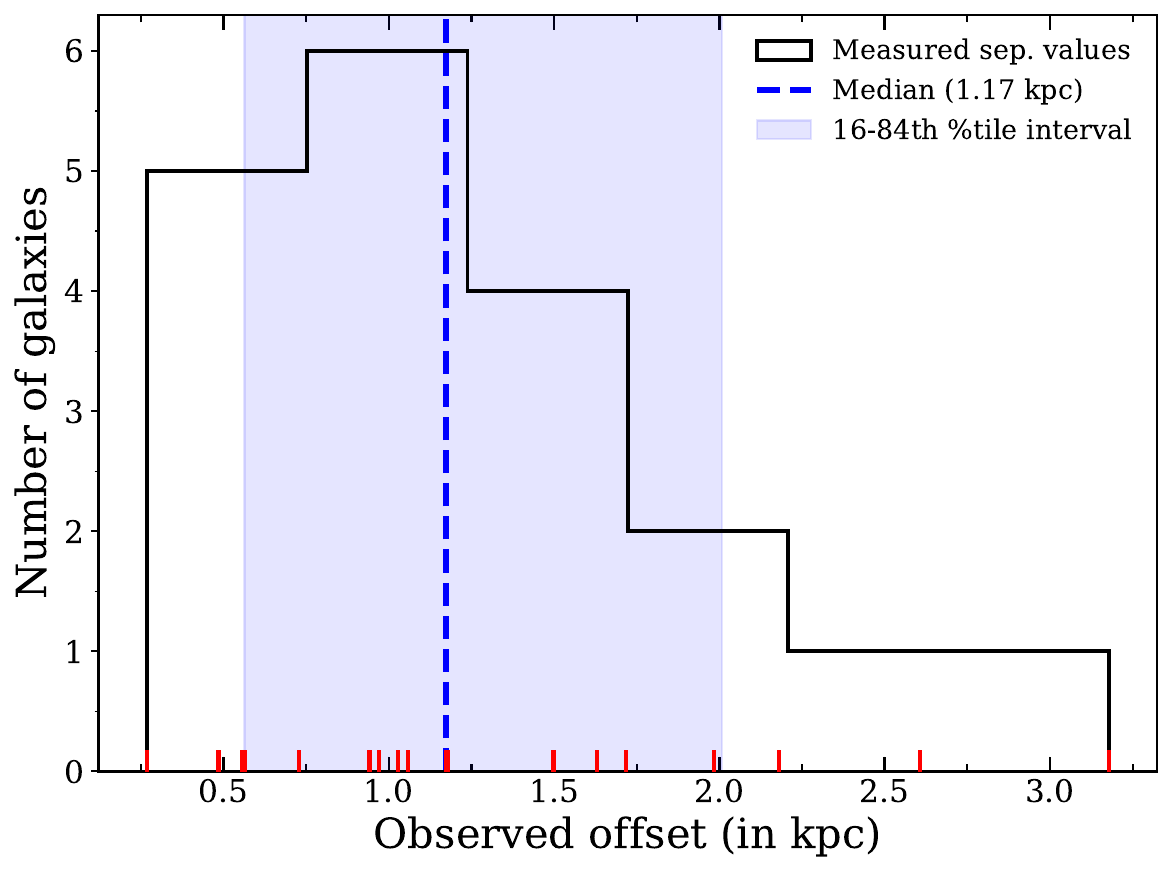}
    \caption{Distribution of the measured HST/F775W–[O{\sc ii}] centroid offsets for the tadpole galaxy sample. The histogram shows the number of galaxies in each offset bin, while the small red rug marks along the bottom indicate the individual measured separations.}
    \label{fig:optical_oii_offset}
\end{figure}

For all 20 sources selected in Sec .~\ref {subsec:sample}, we first determine the MUSE field in which each galaxy is covered. When a source is present in more than one MUSE dataset, we adopt the deepest available cube, following the order MXDF, UDF-10, and HUDF Mosaic, which correspond to decreasing integration times. Based on this criterion, three sources are covered by the MXDF data, while the HUDF Mosaic covers the remaining galaxies. None of the final sample sources fall within the UDF-10 field.

For each galaxy, we use the \texttt{mpdaf} Python package \citep{Bacon_etal2016, Piqueras_etal2017} to extract a cube cutout of size \(6.2^{\prime\prime}\times6.2^{\prime\prime}\), centered on the galaxy position and covering the full MUSE wavelength range. 

To isolate the nebular emission from the stellar continuum, we perform full-spectrum continuum fitting on the MUSE cube. For this purpose, we use the UV-extended E-MILES stellar population synthesis models \citep{Vazdekis_etal2016}, which cover a broad wavelength range of 1680 -- 50000~\text{\AA} at a moderate spectral resolution of \(2.5~\text{\AA}\). The continuum fitting is carried out using \texttt{pPXF} \citep{Cappellari2017}, a full-spectrum fitting code widely used to model galaxy spectra and extract stellar kinematics, gas kinematics, and stellar population properties.

We fit the spectrum of each spaxel across the MUSE wavelength range using the E-MILES templates, allowing us to construct a model of the stellar continuum at every spatial position in the cube. This best-fitting continuum model is then subtracted from the original MUSE cutout, yielding a continuum-subtracted data cube in which the emission-line signal is isolated. These continuum-free cubes are subsequently used to construct [O{\sc ii}] narrow-band images and to extract pseudo-slit spectra for the kinematic measurements.

After producing the continuum-subtracted data cubes, we extract a wavelength-restricted cube centered on the observed [O{\sc ii}] emission for each galaxy, using its MUSE spectroscopic redshift. For this purpose, we select a wavelength window of 100~\text{\AA} around the expected position of the [O{\sc ii}] $\rm \lambda\lambda\ 3726,\ 3729$~\text{\AA} doublet. These [O{\sc ii}] centered cubes are later used for the doublet fitting and kinematic measurements.

To construct the [O{\sc ii}] narrow-band images, we first extract an integrated spectrum from each continuum-subtracted cube using a circular aperture of diameter $\rm \sim 3^{\prime\prime}$ centered on the galaxy. This integrated spectrum is fitted with a double-Gaussian model, in which the two [O{\sc ii}] components are tied to have the same redshift and velocity dispersion. At the same time, their amplitudes are allowed to vary. We further constrain the doublet amplitude ratio, $\rm A_{3727}/A_{3729}$, to lie in the range 0.4 -- 1.5. The best-fitting dispersion, $\sigma$, obtained from the integrated spectrum is then used to define the wavelength range for the narrow-band image. Specifically, we set the lower limit at $2\sigma$ blueward of the first [O{\sc ii}] component and the upper limit at $2\sigma$ redward of the second component. This window encloses the bulk of the [O{\sc ii}] doublet emission ($\rm \sim98\%$) while minimizing the contribution from nearby continuum residuals and noise. Finally, we collapse the continuum-subtracted cube over this wavelength range to produce the [O{\sc ii}] narrow-band image for each galaxy.

\begin{figure*}[!ht]
    \centering
    \includegraphics[width=\linewidth]{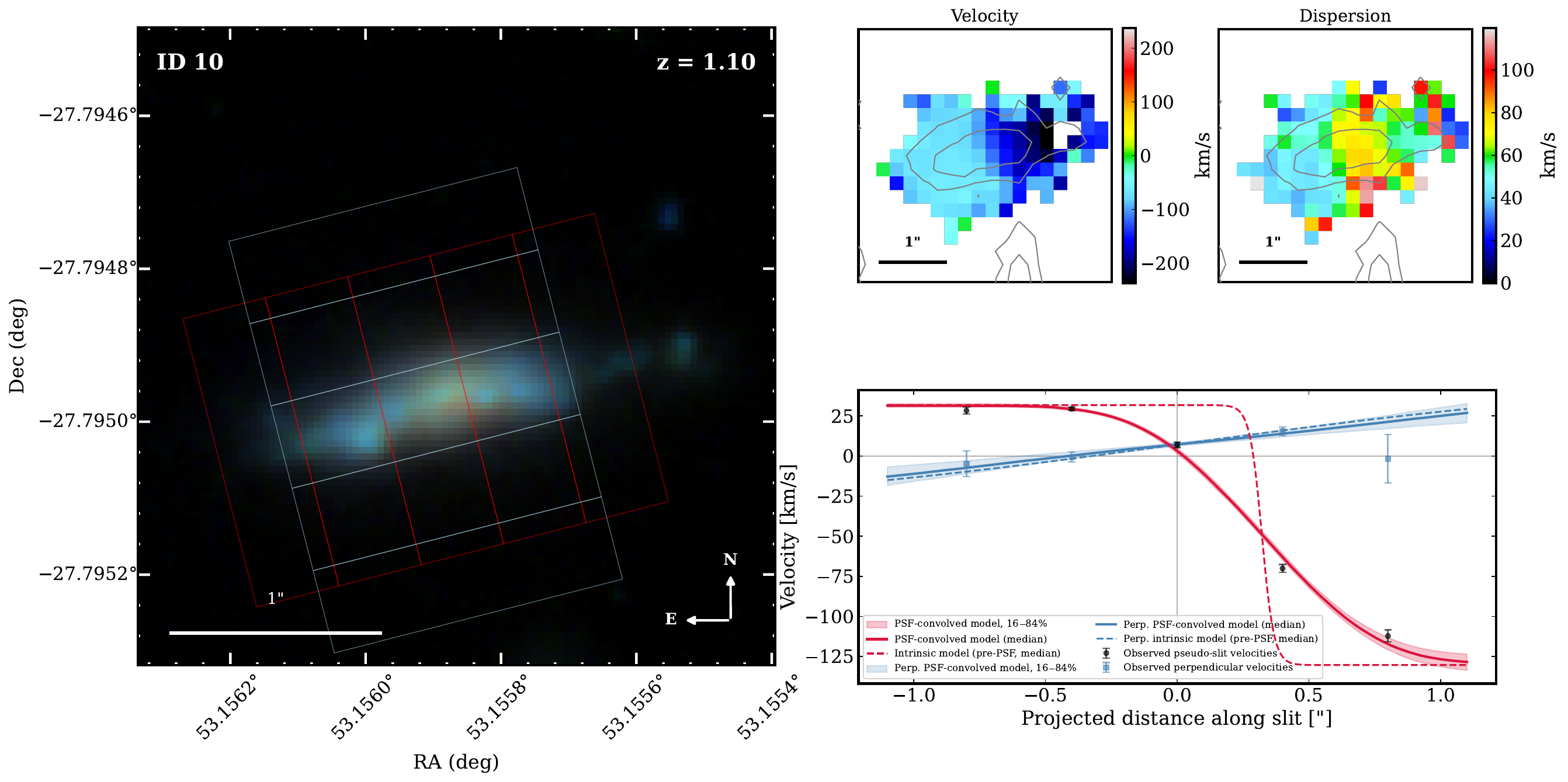} \\
    \includegraphics[width = \linewidth]{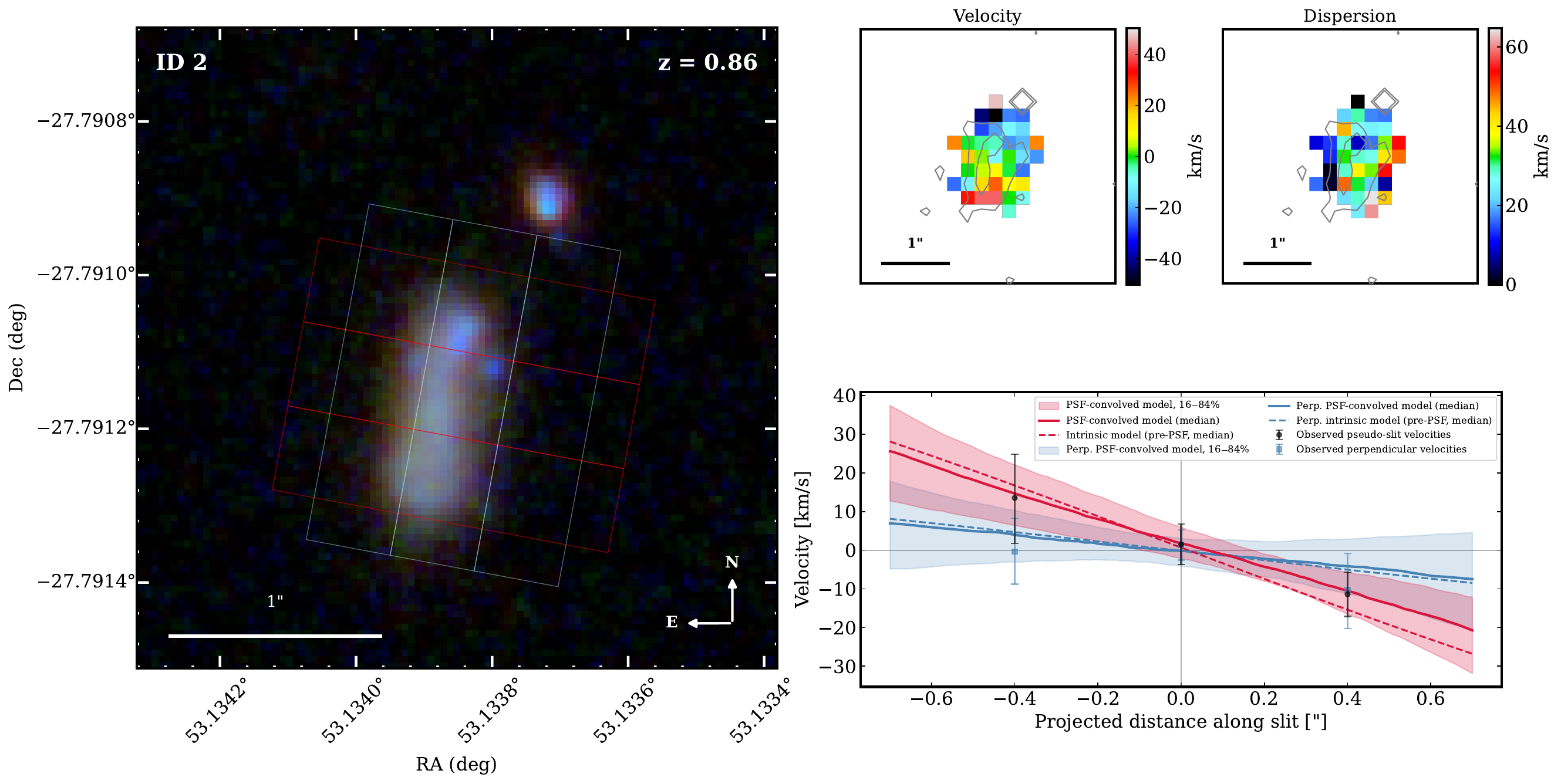}
    \caption{Examples of the spatially resolved [O{\sc ii}] kinematic analysis for two tadpole galaxies (IDs 10 and 2). For each galaxy, the left panel shows the HST/JWST RGB image with the rectangular apertures used to extract the pseudo-slit spectra along the morphological major axis and the perpendicular direction. The upper-right panels show the two-dimensional [O{\sc ii}] line-of-sight velocity and velocity-dispersion maps derived with pPXF, these maps are shown primarily for qualitative comparison because of the limited signal-to-noise of individual spaxels. The lower-right panel shows the velocities measured from the pseudo-slit spectra as a function of projected distance from the [O{\sc ii}] center. Points with error bars indicate the measured velocities, while the solid curves and shaded regions show the median PSF-convolved kinematic models and their 16th–84th percentile intervals for the two slit orientations; the dashed curves show the corresponding intrinsic models before PSF convolution. The $1^{\prime\prime}$ scale bar and north/east directions are indicated in the imaging panels.}
    \label{fig:tpg_kin}
\end{figure*}

\begin{figure*}[ht!]
    \centering
    \addtocounter{figure}{-1}
    \includegraphics[width = \textwidth]{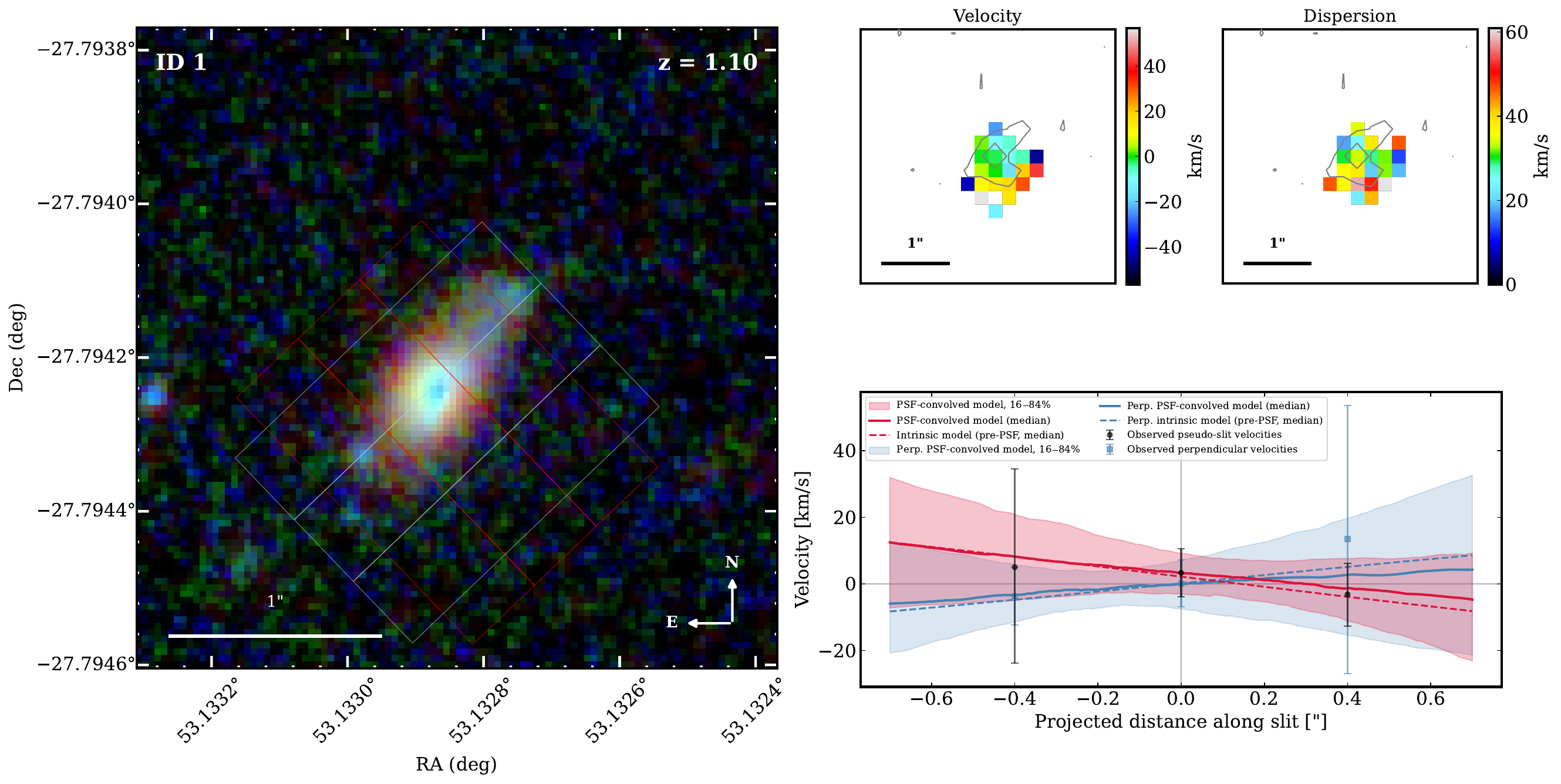} \\
    \includegraphics[width = \textwidth]{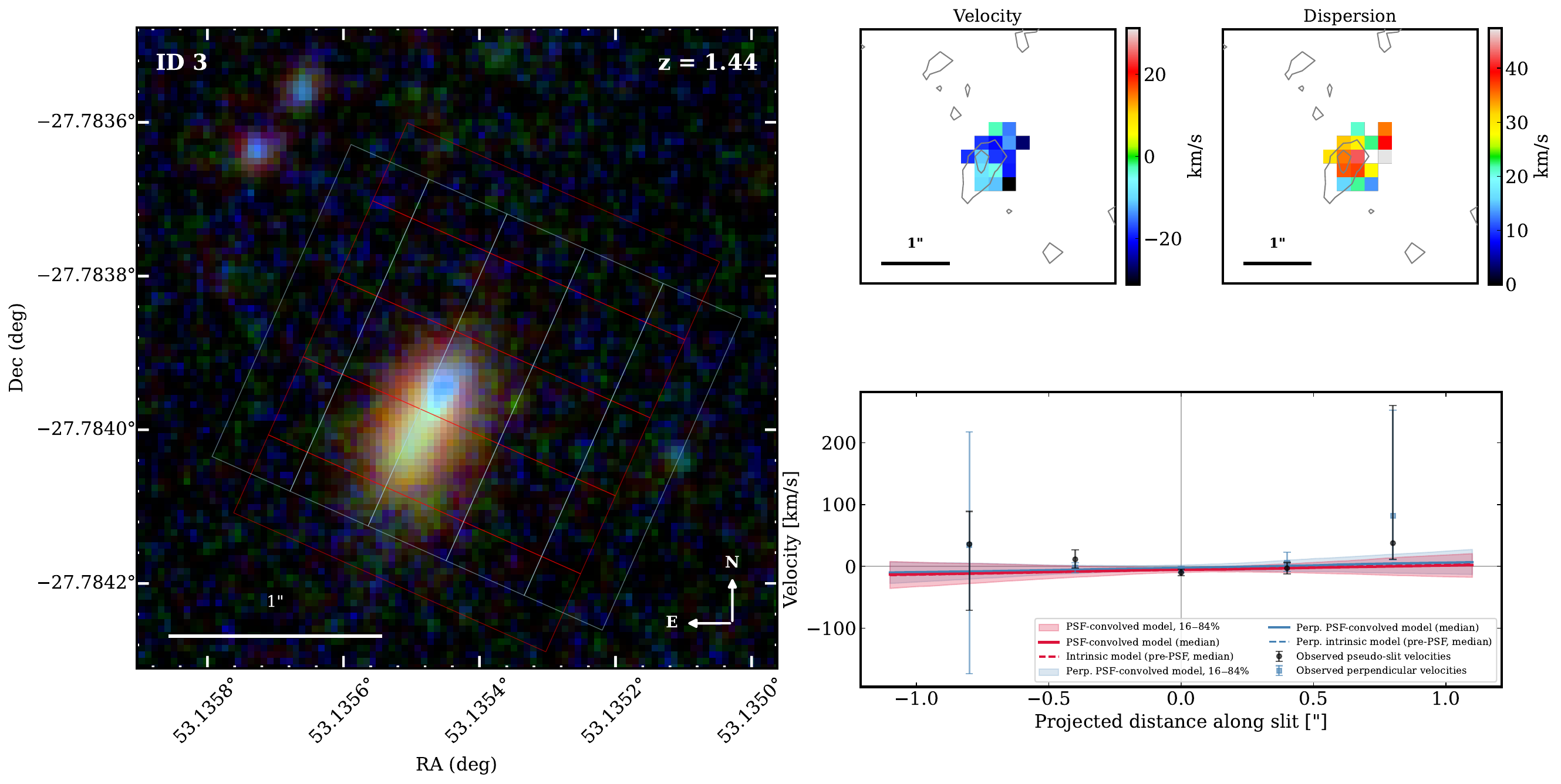}
    \caption{continuation of Figure~\ref{fig:tpg_kin}}\label{fig:tpg_kin_remaining}
\end{figure*}

\begin{figure*}
    \addtocounter{figure}{-1}
    \includegraphics[width = \textwidth]{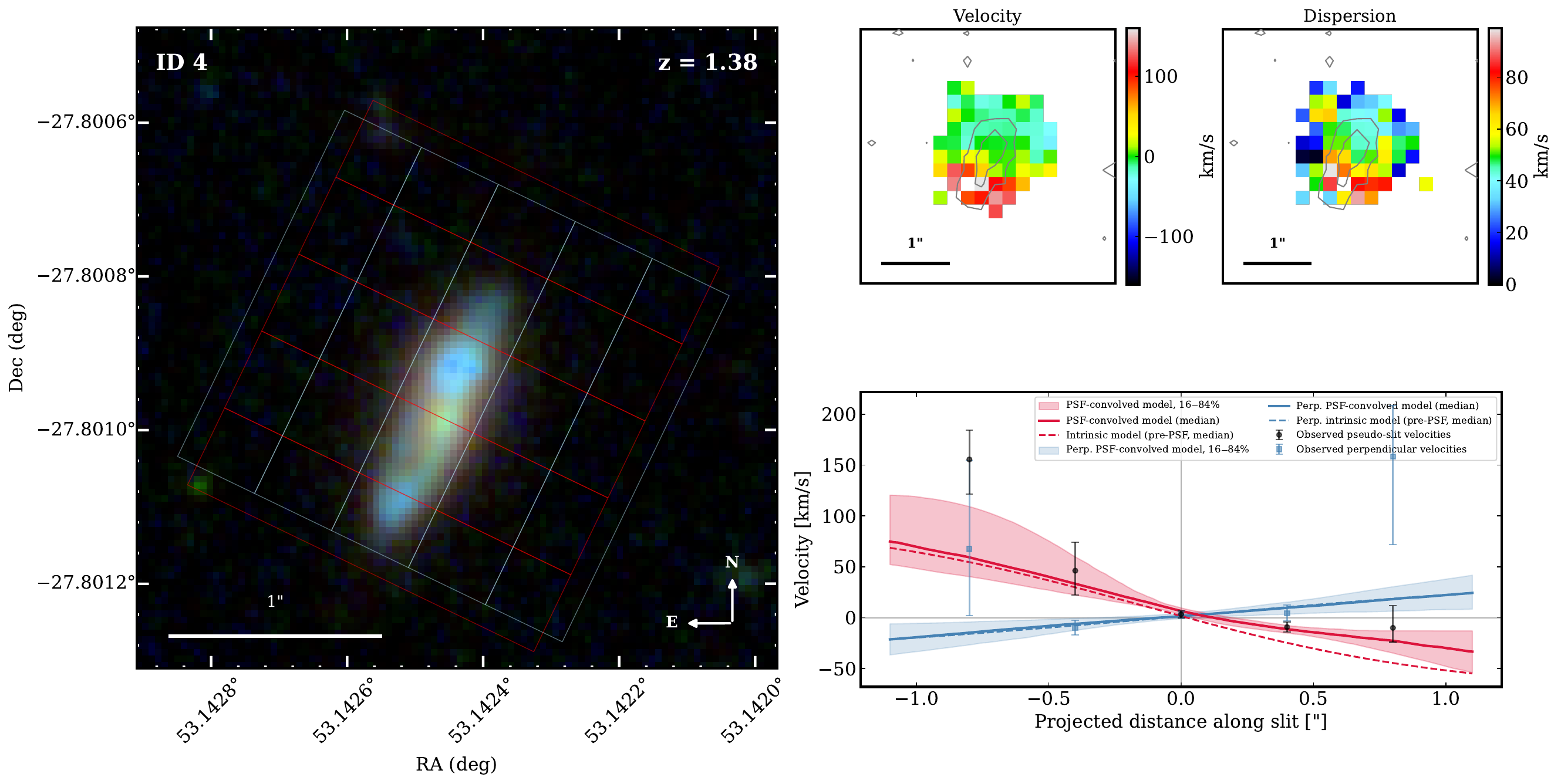} \\
    \includegraphics[width = \textwidth]{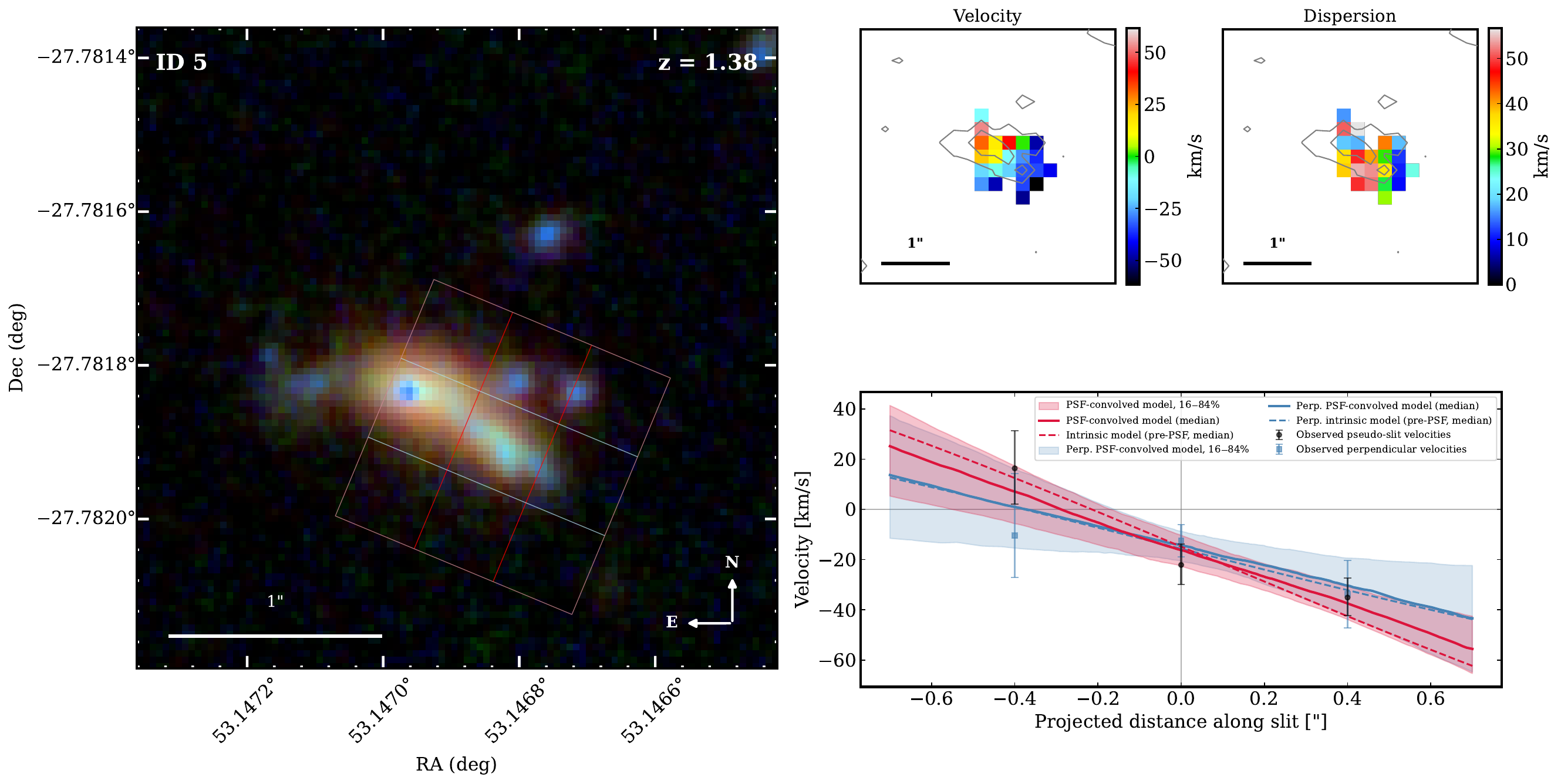}
    \caption{continuation of Figure~\ref{fig:tpg_kin}}
\end{figure*}

\begin{figure*}
    \addtocounter{figure}{-1}
    \includegraphics[width = \textwidth]{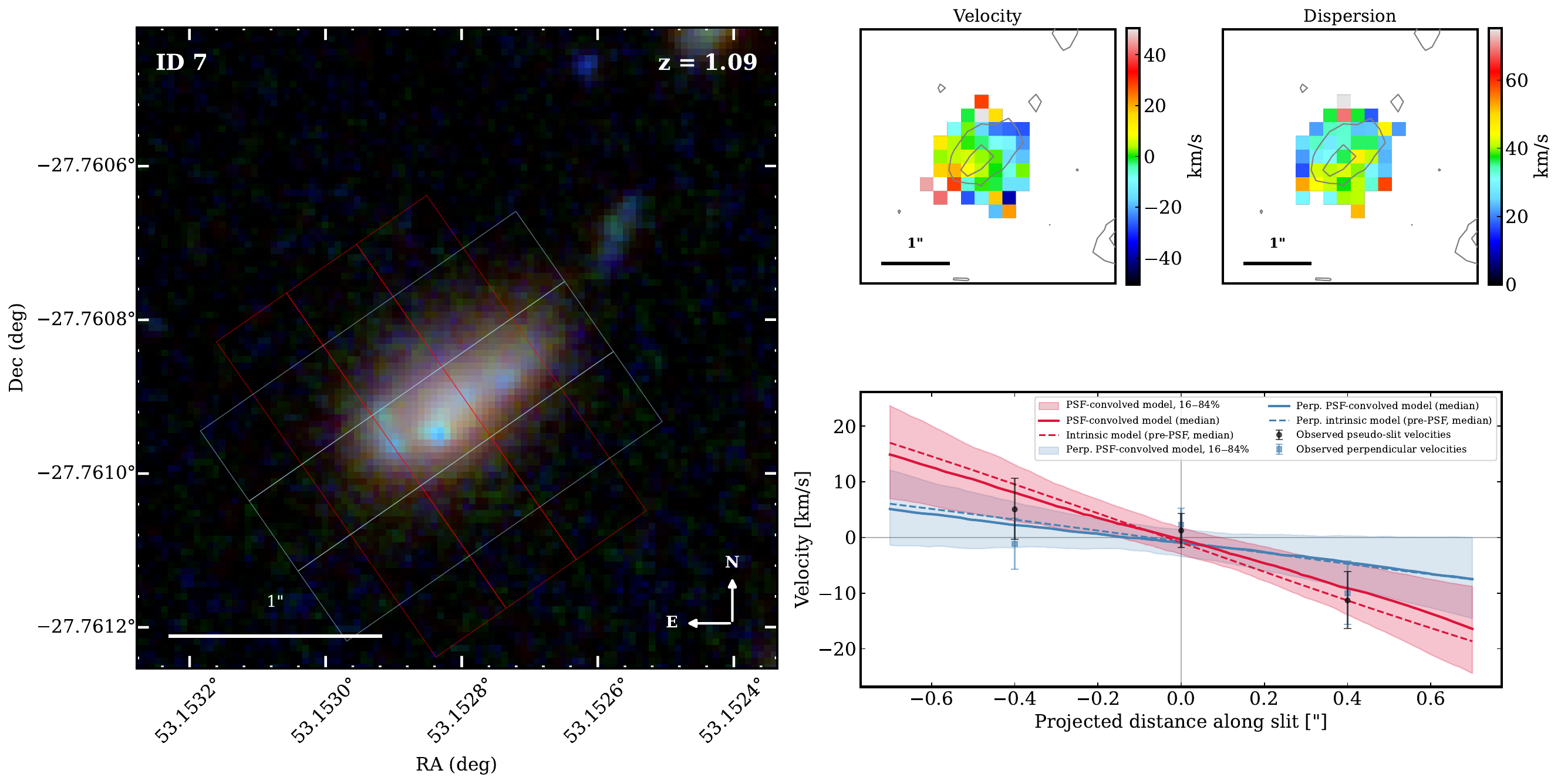}
    \\
    \includegraphics[width = \textwidth]{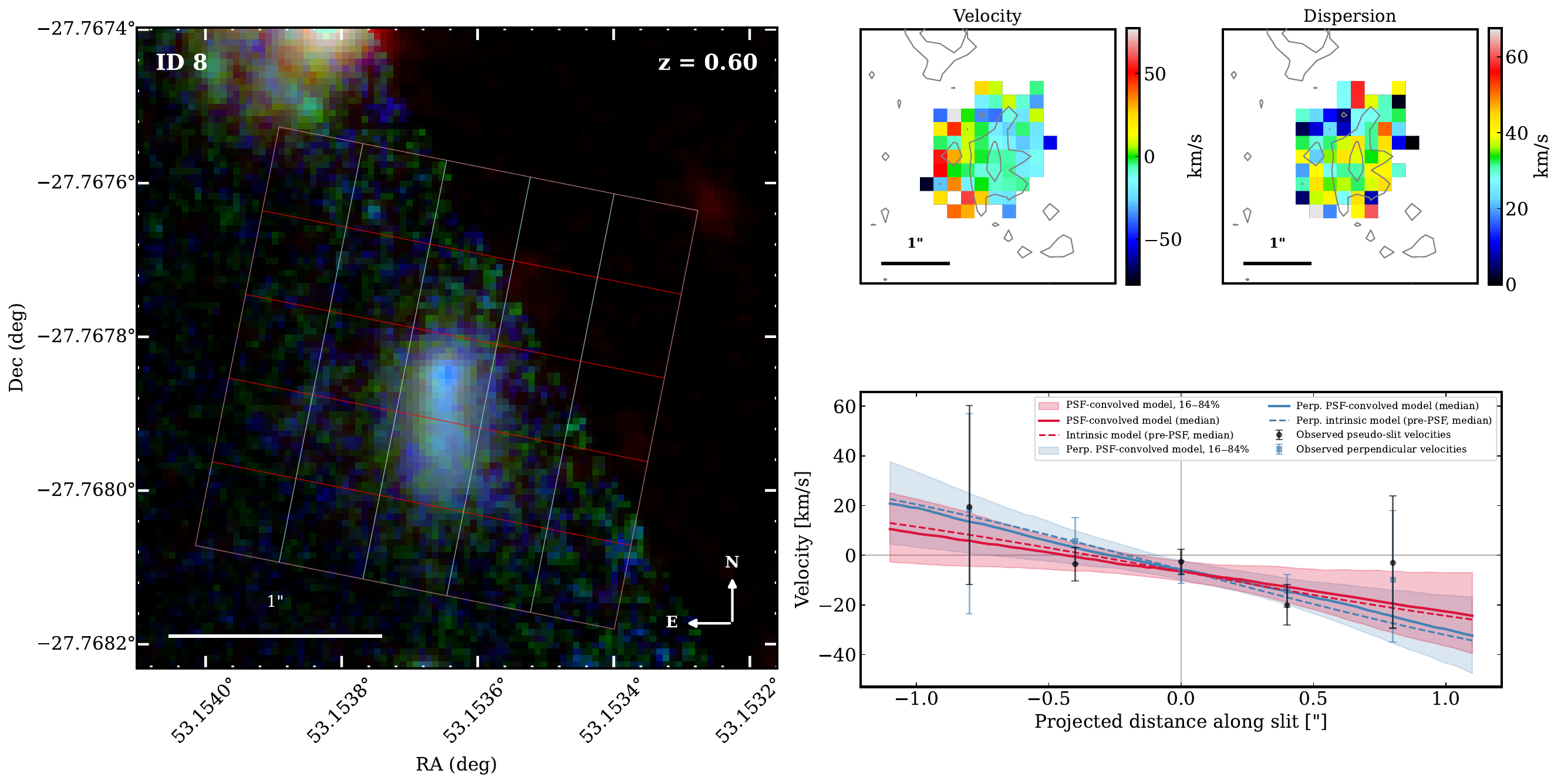}
    \caption{continuation of Figure~\ref{fig:tpg_kin}}
\end{figure*}

\begin{figure*}
    \addtocounter{figure}{-1}
    \includegraphics[width = \textwidth]{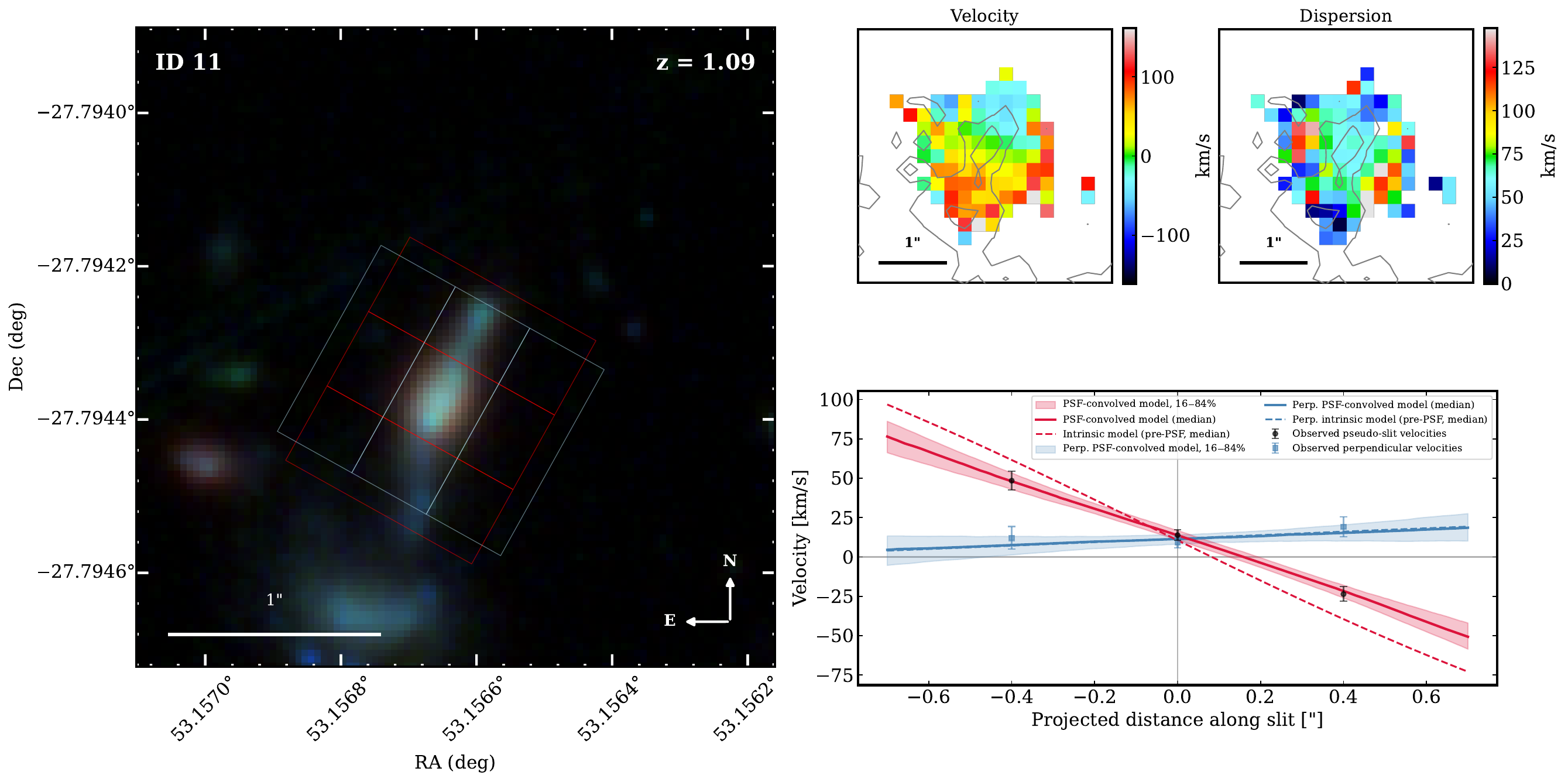} \\
    \includegraphics[width = \textwidth]{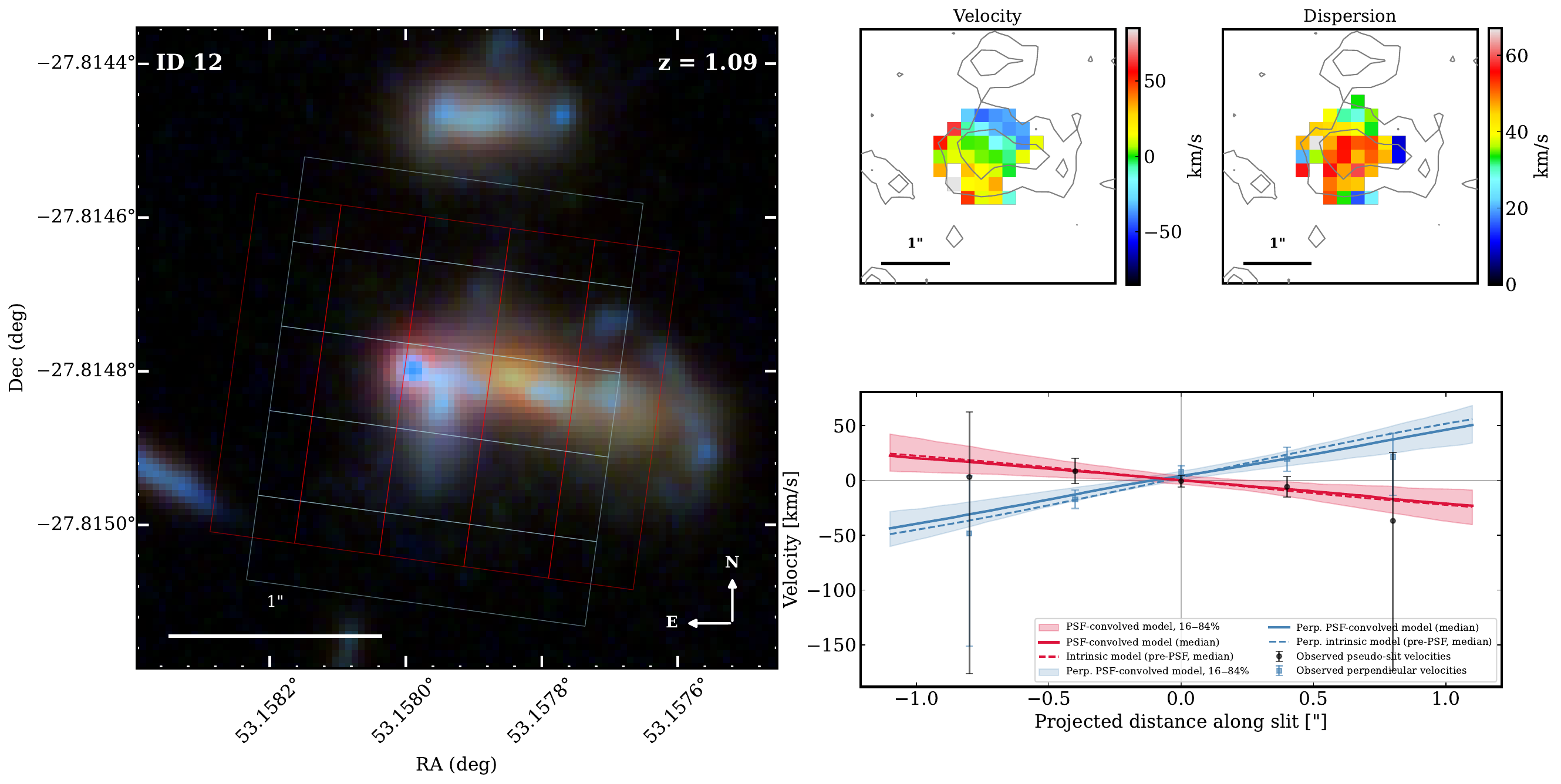}
    \caption{continuation of Figure~\ref{fig:tpg_kin}}
\end{figure*}

\begin{figure*}
    \addtocounter{figure}{-1}
    \includegraphics[width = \textwidth]{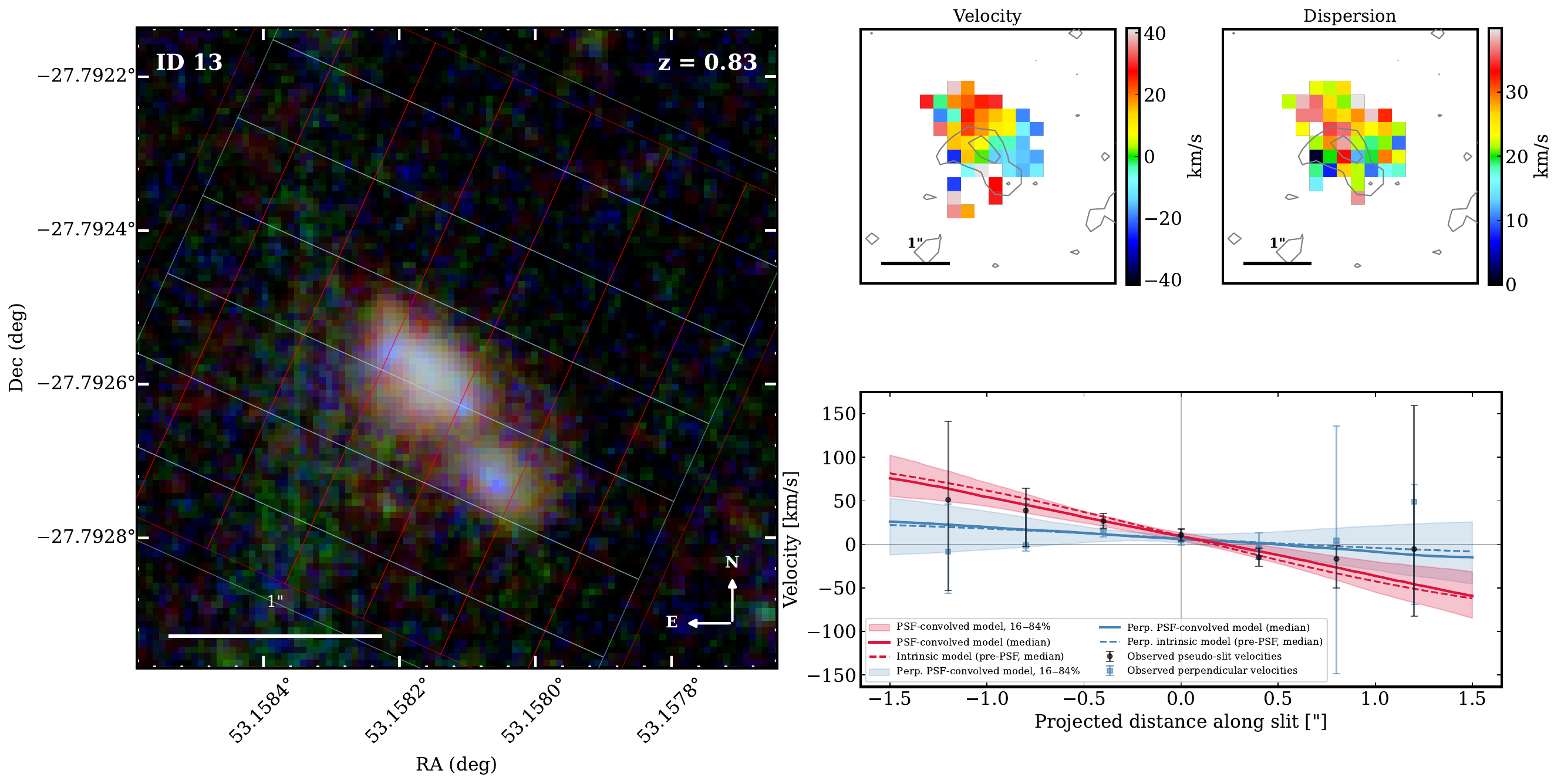} \\
    \includegraphics[width = \textwidth]{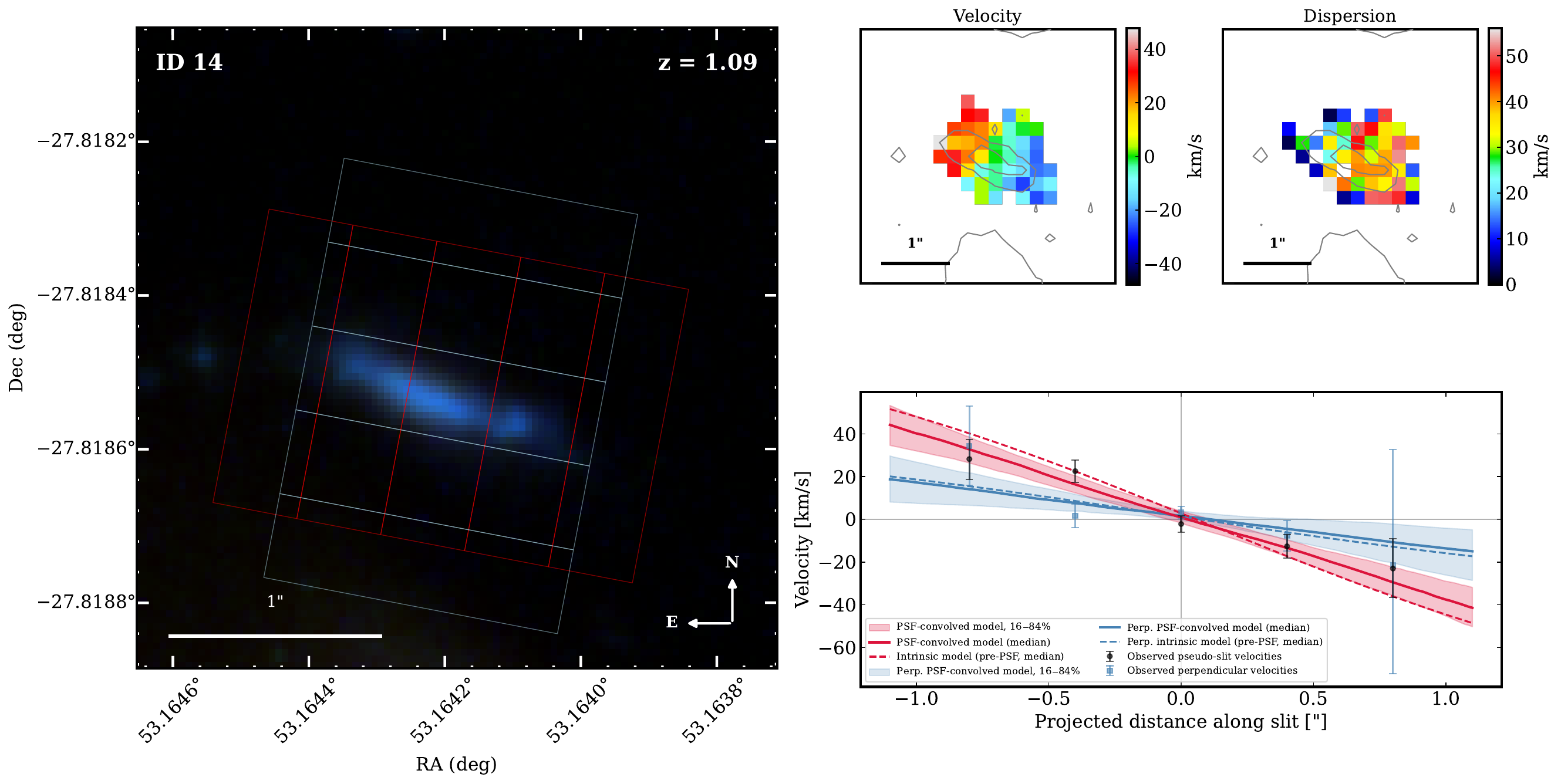}
    \caption{continuation of Figure~\ref{fig:tpg_kin}}
\end{figure*}

\begin{figure*}
    \addtocounter{figure}{-1}
    \includegraphics[width = \textwidth]{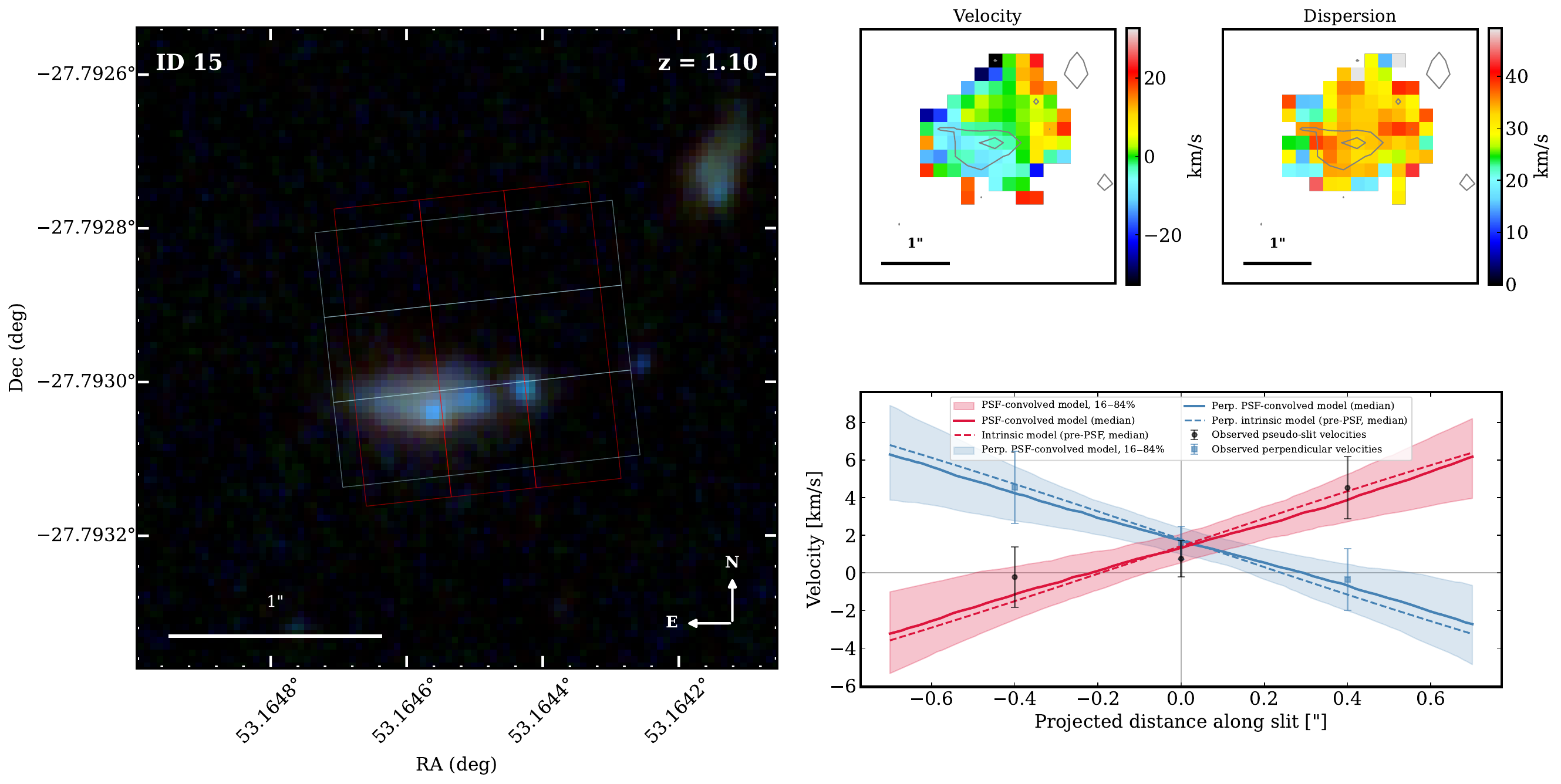} \\
    \includegraphics[width = \textwidth]{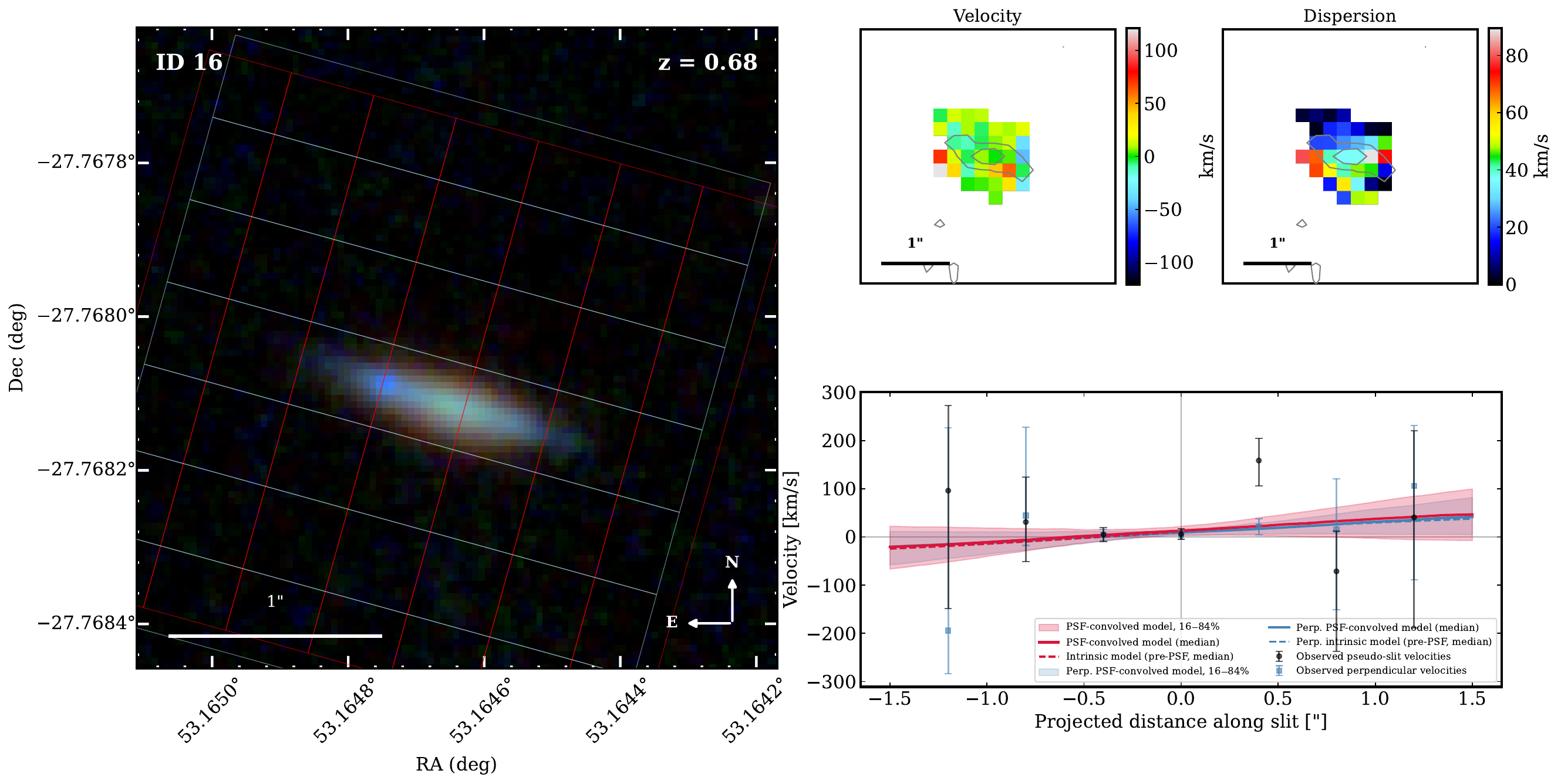}
    \caption{continuation of Figure~\ref{fig:tpg_kin}}
\end{figure*}

\begin{figure*}
    \addtocounter{figure}{-1}
    \includegraphics[width = \textwidth]{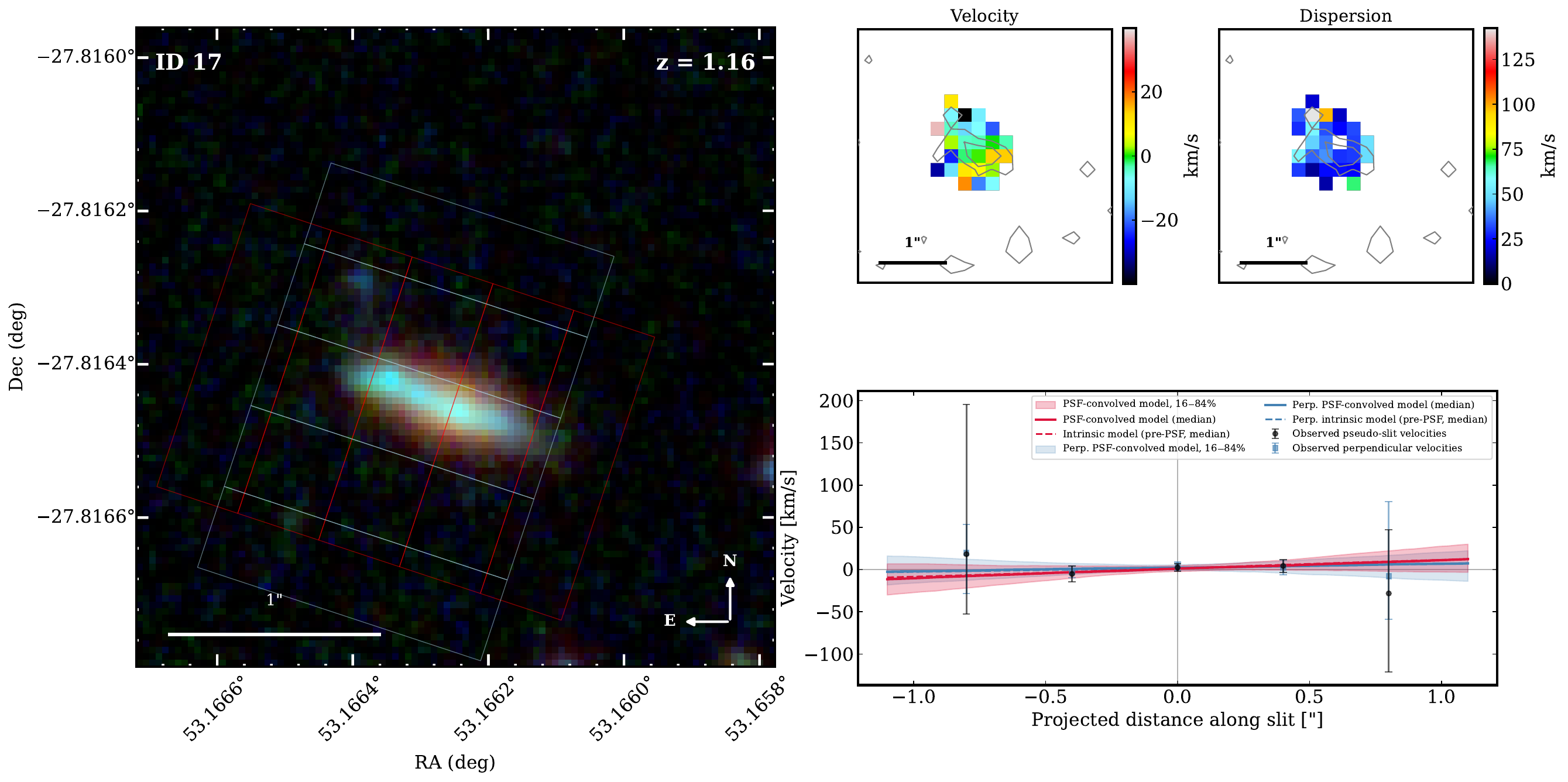} \\
    \includegraphics[width = \textwidth]{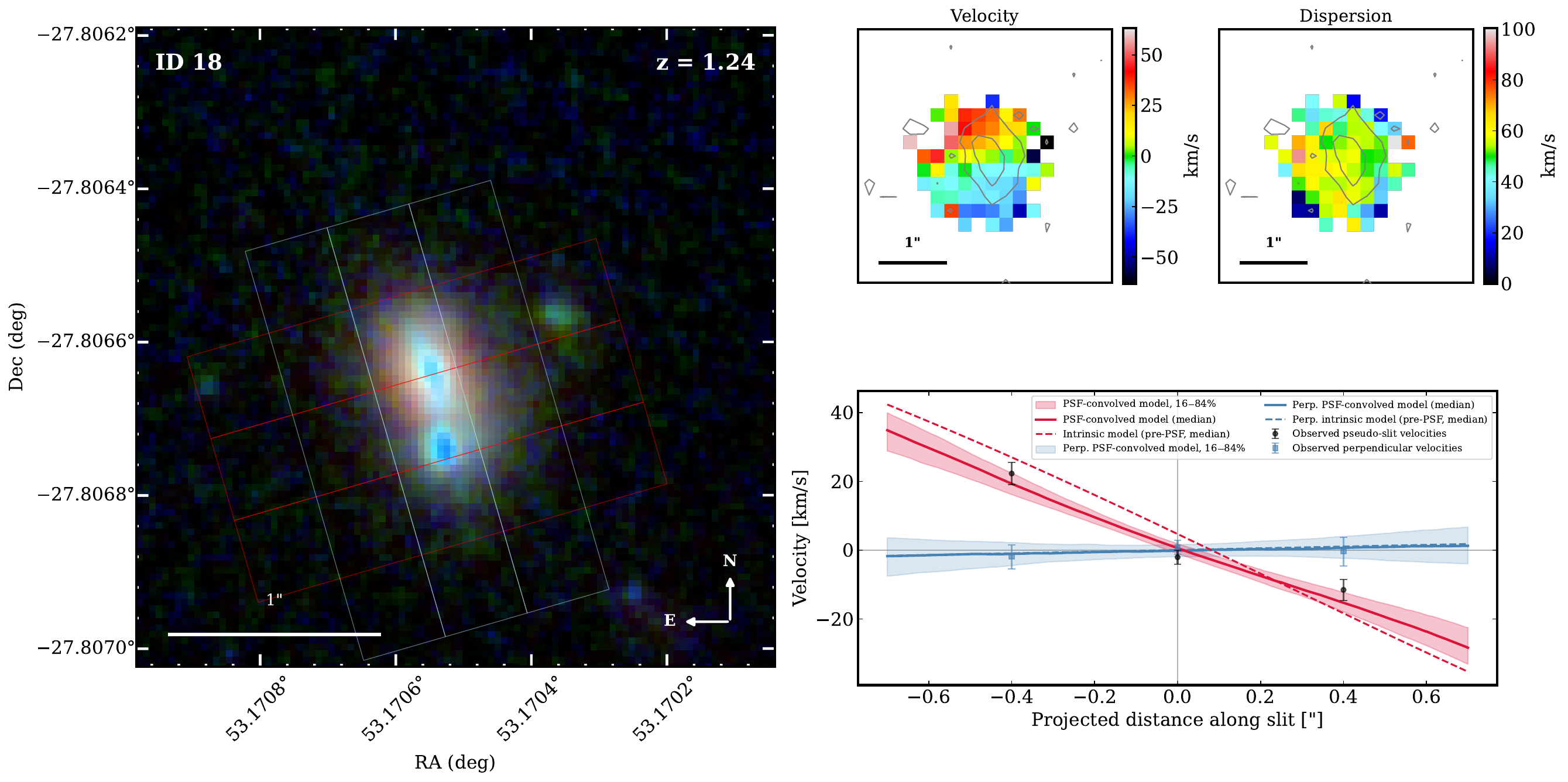}
    \caption{continuation of Figure~\ref{fig:tpg_kin}}
\end{figure*}

\begin{figure*}
    \addtocounter{figure}{-1}
    \includegraphics[width = \textwidth]{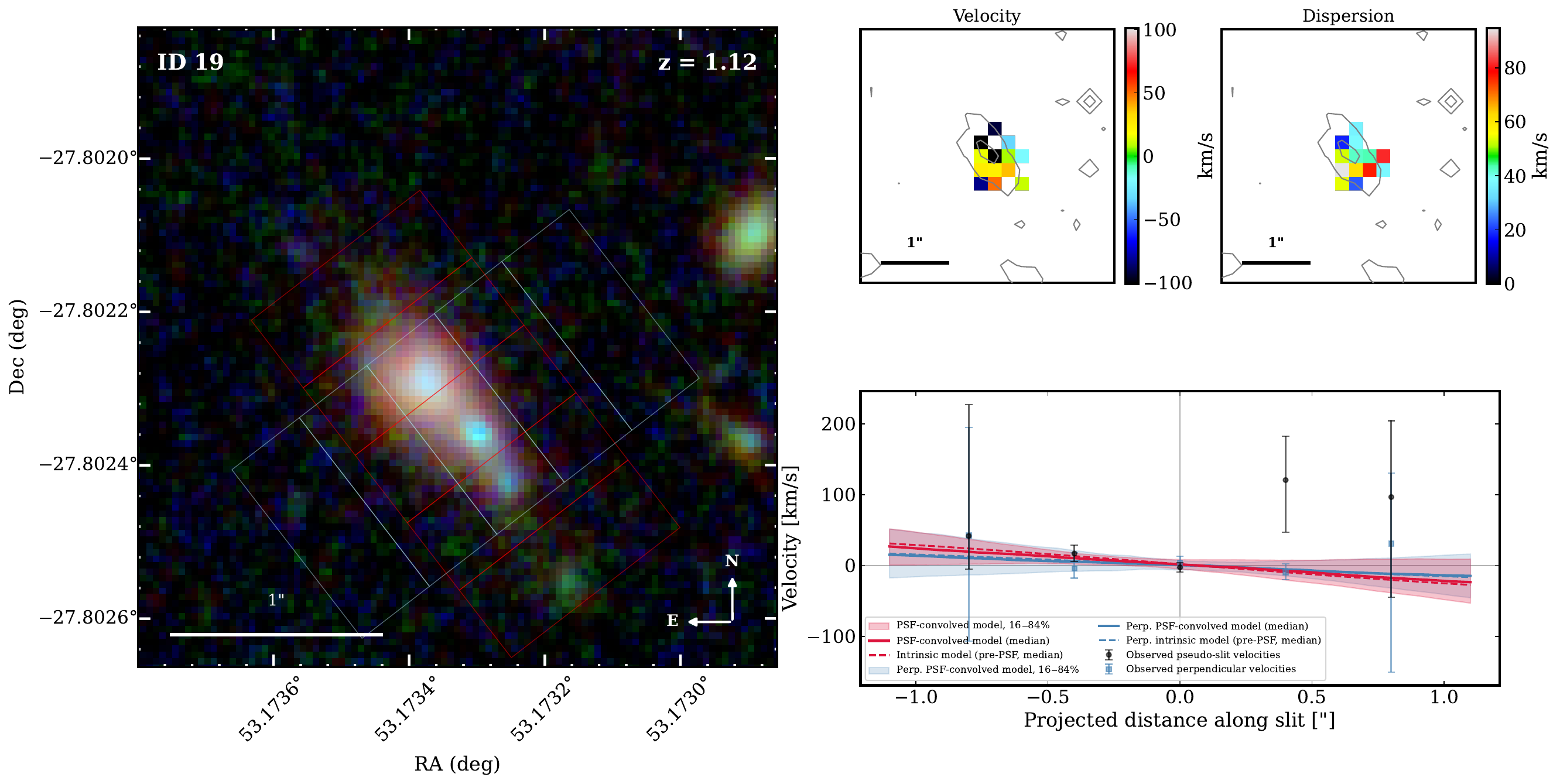} \\
    \includegraphics[width = \textwidth]{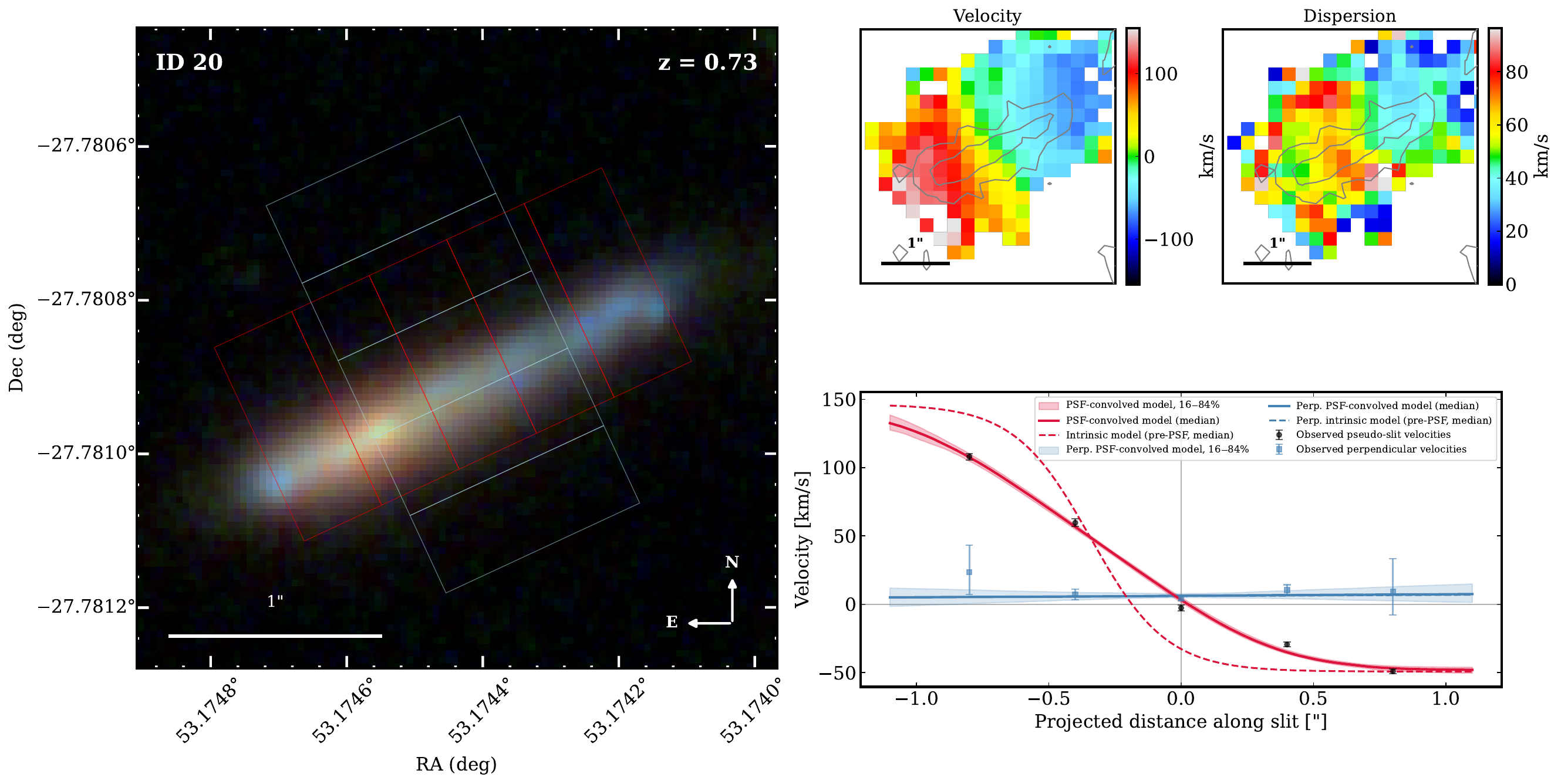}
    \caption{continuation of Figure~\ref{fig:tpg_kin}}
\end{figure*}

\begin{table*}[!ht]
    \centering
    \hspace*{-3.7cm}
    \renewcommand{\arraystretch}{1.15}
    \begin{tabular}{|c|c|c|c|c|c|c|c|c|c|c|c|}
    \hline
        ID & RA  & Dec  & z & $\rm \Delta v_{major}$ & $\rm \Delta v_{minor}$ & $\rm PA_{morph}$ & $\rm PA_{kin}$ & $\rm log M_{*}$ & $\rm log SFR$ & $\rm \Delta MS$ & d\\
           & (deg.) & (deg.) & & $\rm (kms^{-1})$ & $\rm (kms^{-1})$ & (deg) & (deg) & $\rm (M_{\odot})$ & $(\rm M_{\odot}yr^{-1})$ & (dex) & (kpc)\\
    \hline
         1  & 53.13283920 & -27.79418754 & 1.0974 & $8.31_{-30.32}^{+30.96}$ & $17.14_{-41.24}^{+40.93}$ & -- & -- & 8.182 & -0.466 & +0.400 & 1.72\\
         2  & 53.13385167 & -27.79109476 & 0.8588 & $24.9_{-13.05}^{+12.69}$ & $10.09_{-12.81}^{+13.05}$ & 169 & $150_{-4}^{+4}$ & 8.373 & -0.475 & +0.323 & 1.03\\
         3  & 53.13543319 & -27.78389358 & 1.4379 & $49.67_{-97.78}^{+252.45}$ & $85.91_{-215.02}^{+262.05}$ & -- & -- & 8.789 & -0.047 & +0.201 & 2.18\\
         4  & 53.14243852 & -27.80089314 & 1.3818 & $162.01_{-39.34}^{+33.52}$ & $75.84_{-118.80}^{+90.94}$ & 154 & $175_{-3}^{+3}$ & 8.925 & 0.434 & +0.587 & 1.17\\
         5  & 53.14689151 & -27.78177745 & 1.3824 & $51.34_{-16.25}^{+16.75}$ & $24.49_{-22.10}^{+27.80}$ & 67 & $62_{-6}^{+6}$ & 8.986 & 0.553 & +0.655 & 3.18\\
         6  & 53.15124409 & -27.78947857 & 0.4238 & -- & -- & -- & -- & 7.199 & -1.854 & +0.120 & 1.50\\
         7  & 53.15280695 & -27.76083663 & 1.0932 & $16.38_{-7.46}^{+7.51}$ & $8.85_{-7.31}^{+7.28}$ & 125 & $117_{-4}^{+4}$ & 8.619 & -0.057 & +0.447 & 0.94\\
         8  & 53.15363044 & -27.76781451 & 0.6049 & $23.99_{-42.11}^{+48.10}$ & $26.7_{-49.77}^{+46.66}$ & 169 & $110_{-7}^{+7}$ & 8.286 & -0.701 & +0.281 & 0.73\\
         9  & 53.15562438 & -27.76048278 & 1.2221 & $55.98_{-62.30}^{+66.35}$ & $19.39_{-44.88}^{+46.17}$ & -- & -- & 8.161 & -0.565 & +0.273 & --\\
         10 & 53.15586562 & -27.79490145 & 1.0959 & $140.50_{-4.38}^{+4.38}$ & $3.09_{-17.20}^{+17.02}$ & 104 & 1$09_{-0.4}^{+0.4}$ & 9.410 & 0.949 & +0.796 & 0.48\\
         11 & 53.15663082 & -27.79430552 & 1.0945 & $72.01_{-7.51}^{+7.52}$ & $7.07_{-9.66}^{+9.54}$ & 151 & $157_{-2}^{+2}$ & -- & -- & -- & 0.27\\
         12 & 53.15792465 & -27.81476974 & 1.0937 & $37.9_{-202.94}^{+161.22}$ & $67.01_{-63.86}^{+106.84}$ & 82 & $154_{-3}^{+3}$ & -- & -- & -- & 1.98\\
         13 & 53.15811539 & -27.79255295 & 0.8325 & $35.67_{-191.83}^{+131.93}$ & $39.91_{-133.77}^{+63.66}$ & -- & -- & 8.486 & -0.682 & +0.033 & 1.06\\
         14 & 53.16418209 & -27.81846837 & 1.0946 & $51.08_{-16.81}^{+16.41}$ & $55.83_{-57.20}^{+54.08}$ & 79 & $63_{-2}^{+2}$ & 9.000 & -0.287 & -0.100 & 1.18\\
         15 & 53.16452604 & -27.79295578 & 1.0952 & $4.75_{-2.30}^{+2.31}$ & $4.91_{-2.52}^{+2.53}$ & -- & -- & 8.140 & -0.340 & +0.561 & 2.61\\
         16 & 53.16464004 & -27.76804090 & 0.6773 & $51.99_{-305.08}^{+293.81}$ & $229.82_{-468.53}^{+207.93}$ & 74 & $66_{-16}^{+16}$ & 8.061 & -1.229 & -0.094 & 0.56\\
         17 & 53.16624832 & -27.81637764 & 1.1637 & $68.63_{-118.46}^{+198.10}$ & $17.25_{-98.44}^{+67.17}$ & 71 & $37_{-16}^{+16}$ & 8.450 & -0.358 & +0.262 & 0.97\\
         18 & 53.17051168 & -27.80660654 & 1.2445 & $33.83_{-4.51}^{+4.56}$ & $1.5_{-5.42}^{+5.39}$ & 16 & $15_{-2}^{+2}$ & 9.121 & 0.145 & +0.180 & 0.56\\
         19 & 53.17332847 & -27.80224607 & 1.1234 & $12.18_{-231.58}^{+231.58}$ & $33.53_{-191.42}^{+232.54}$ & -- & -- & 8.519 & -0.323 & +0.253 & 1.50\\
         20 & 53.17443783 & -27.78086223 & 0.7341 & $157.09_{-2.84}^{+2.82}$ & $13.64_{-29.03}^{+26.50}$ & --& -- & 9.076 & -0.082 & +0.185 & 1.63\\
    \hline
    \end{tabular}
    
    \caption{Properties and kinematic measurements of the final tadpole-galaxy sample. Columns list the galaxy ID, right ascension and declination, MUSE spectroscopic redshift, edge-to-edge [O\,{\sc ii}] velocity shear measured along the F775W morphological major axis ($\Delta v_{\rm major}$) and along the perpendicular direction ($\Delta v_{minor}$), morphological position angle ($PA_{\rm morph}$) measured from the HST F775W image as $z\sim1$ it observes near-UV optical emission which traces both young and intermediate stellar population which is appropriate for tadpole like star-forming galaxies, kinematic position angle ($PA_{\rm kin}$) derived from the direction of the maximum projected [O\,{\sc ii}] velocity gradient, projected physical separation $d$ between the F775W and [O\,{\sc ii}] centroids, stellar mass, star-formation rate, and offset from the star-forming main sequence, $\Delta_{\rm MS}$. Uncertainties correspond to the quoted 68\% confidence intervals. Dashes indicate cases for which a reliable measurement could not be obtained.}
    \label{tab:placeholder}
\end{table*}

\section{Morphological measurements and HST/F775W -- [O{\sc ii}] offset} \label{sec:optical_o2_offset}

We measure the parameters of galaxy morphology, mainly the centroid and PA, using the \texttt{find\_galaxy()} method of the \texttt{mgefit} Python module \cite{Cappellari2002}. This method determines the luminosity-weighted center and position angle from the intensity distribution of the selected image pixels. It follows the standard convention for measuring PA and measures the angle relative to the +Y direction, assuming the +Y direction is aligned with equatorial north, which is the case for the HST images. We then measure the centers of the [O{\sc ii}] line-only emission from the narrow-band images (see Sec.\ref{sec:cube_contisub_nb}) and calculate the separation between the two in kpc, with their distribution shown in Figure~\ref{fig:optical_oii_offset}. In this analysis, only ID 9 galaxy is not considered as its [O{\sc ii}] emission is not detected in the narrowband imaging, hence we don't see any contours on the RGB image ID9 in the Figure~\ref{fig:tpg_kin_rgb}. 

Before measuring the morphological parameters, we visually inspected
the F775W cutouts for contamination from unrelated neighboring
sources. Since the luminosity-weighted moments used by
\texttt{find\_galaxy()} can be biased by nearby objects, contaminating
sources were masked prior to the final centroid and position-angle
measurements. The morphological centers and position angles used
throughout the subsequent analysis were then remeasured from these
cleaned F775W images. We use a segmentation map to isolate the target and replace pixels from neighboring sources with background-representative values within the cutout, assuming a Gaussian distribution.

For the 19 galaxies with measurable F775W--[O\,{\sc ii}] centroids,
16 ($\sim84\%$) have projected separations smaller than $2$ kpc,
while 18 ($\sim95\%$) lie within $3$ kpc. Only one galaxy has a
separation larger than $3$ kpc, with the maximum measured offset being
$\sim3.2$ kpc. The F775W and [O\,{\sc ii}] emission are therefore
spatially well aligned in the majority of the sample, although a few
systems show appreciable displacements.

\section{Pseudo-slit extraction} \label{sec:Pseudo_slit_extraction}

Although MUSE provides spatially resolved spectroscopy, most galaxies in our sample are faint. In several cases, spaxel-by-spaxel or uniformly binned velocity maps are too noisy to provide reliable kinematic information. We therefore adopt a pseudo-slit approach, which increases the signal-to-noise ratio while preserving spatial information along physically motivated directions, i.e., along the photometric major and minor axes.

To position the central pseudo-slit, we use the [O{\sc ii}] narrow-band image, since [O{\sc ii}] is used as a tracer of ionized-gas kinematics. We determine the center and basic structural parameters of the [O{\sc ii}] emission using the \texttt{find\_galaxy()} method, which is applied to the narrow-band image constructed as described in Sec .~\ref {sec:cube_contisub_nb}. The measured semi-major and semi-minor axes are then used to define the spatial extent of the slit system, including the slit length and the number of extraction apertures placed across each galaxy.

The orientation of the primary pseudo-slit is fixed using the photometric position angle measured from the HST F775W image. This slit is aligned with the galaxy's morphological major axis and is used to trace projected velocity variations along its elongated structure. In addition, we place a second set of pseudo-slits perpendicular to the photometric major axis. The velocity gradients measured along the major and perpendicular directions are then compared to examine whether the observed gas kinematics are consistent with ordered rotation or show signatures of disturbed or non-circular motions. We also fixed the width of each slit to 2 pixels ($\rm 0.4^{\prime\prime}$ with MUSE pixel scale). 

The spectrum corresponding to each pseudo-slit aperture is extracted using the \texttt{Photutils} package. For every wavelength slice of the continuum-subtracted cube, we sum the flux from all spatial pixels enclosed within the rectangular aperture. Repeating this procedure over the desired wavelength range around [O{\sc ii}] produces a spectrum for each slit position. The corresponding variance spectrum is extracted from the variance cube using the same aperture geometry. Since the flux is summed over the enclosed pixels, the individual pixel variances are also summed, and the square root of the resulting variance spectrum is taken as the uncertainty on the extracted flux. This ensures that the spectral errors are propagated consistently for each pseudo-slit.

\section{[O{\sc ii}] Kinematics from MUSE data} \label{sec:o2_kin_muse}

To calculate the velocities from the extracted pseudo slit spectra, we use the same double Gaussian fitting as used in the Sec. \ref{sec:cube_contisub_nb} to fit the line in each slit and then convert into measured velocity. The double Gaussian is fixed at the observed wavelength with redshift of the galaxy, which we remeasure using the pPXF module, and the additional free velocity parameter is included in the fitting as described in \ref{eqn:velocity_measure}. The line-fitting was performed using the MCMC approach, with the Python \texttt{emcee} \citep{Foreman-Mackey2013, Foreman-Mackey2019} module to account for errors, especially in the velocity.

\begin{equation}
    \lambda_{obs} = \lambda_{rest}(1 + z)e^{v/c}
    \label{eqn:velocity_measure}
\end{equation}

In each velocity measurement, the median of the marginalized posterior is taken as the velocity, and the 16th and 84th percentiles are treated as the errors on the measurement.

\subsection{Two-dimensional kinematic maps} \label{subsec:2d_kin_maps}

To obtain a qualitative view of the spatially resolved gas kinematics, we construct two-dimensional line-of-sight velocity and velocity-dispersion maps for each galaxy using the MUSE datacubes. The spectrum in each individual spaxel is fitted with \textsc{pPXF}, including the nebular emission lines available within the MUSE wavelength range, and the resulting gas velocity and velocity dispersion are used to construct the corresponding two-dimensional maps. The velocities are measured relative to the systemic redshift of each galaxy.

Examples of the resulting maps are shown in Figure~\ref{fig:tpg_kin}. The velocity maps provide a visual indication of the direction and amplitude of the spatial velocity variation, while the velocity-dispersion maps show the corresponding distribution of the observed line width. In some galaxies, a coherent velocity gradient is visible across the source, whereas in others the velocity field is weak or irregular. However, the galaxies in our sample are generally faint and compact, and the emission-line signal-to-noise ratio is insufficient in many individual MUSE spaxels to obtain robust kinematic measurements across the full spatial extent of the galaxy. Consequently, the two-dimensional maps can contain large spaxel-to-spaxel variations and are sensitive to the limited spatial resolution and signal-to-noise of the observations.

Because several galaxies are only marginally resolved with MUSE, the observed pseudo-slit velocity profiles can be affected by beam smearing. We therefore compare the measured velocities with a simple PSF-convolved kinematic model. For each slit orientation, an intrinsic line-of-sight velocity profile is generated and convolved with the MUSE PSF before being sampled at the same projected positions as the observations.

The model parameters are constrained using MCMC, accounting for the asymmetric velocity uncertainties. The posterior median is adopted as the representative model, while the 16th--84th percentile range is used to show the uncertainty. In Figure~\ref{fig:tpg_kin}, dashed curves show the intrinsic profiles, whereas the solid curves and shaded regions show the corresponding PSF-convolved models and their uncertainties.

Given the limited spatial resolution and large uncertainties for some galaxies, we use these models primarily to assess the effects of beam smearing and to determine whether the observed velocities are consistent with a coherent spatial gradient, rather than to derive precise circular velocities.

\subsection{Velocity shear along the major/minor axis} \label{subsec:delv}

\begin{figure*}[!ht]
    \centering
    \subfigure{\includegraphics[width = 0.7\textwidth]{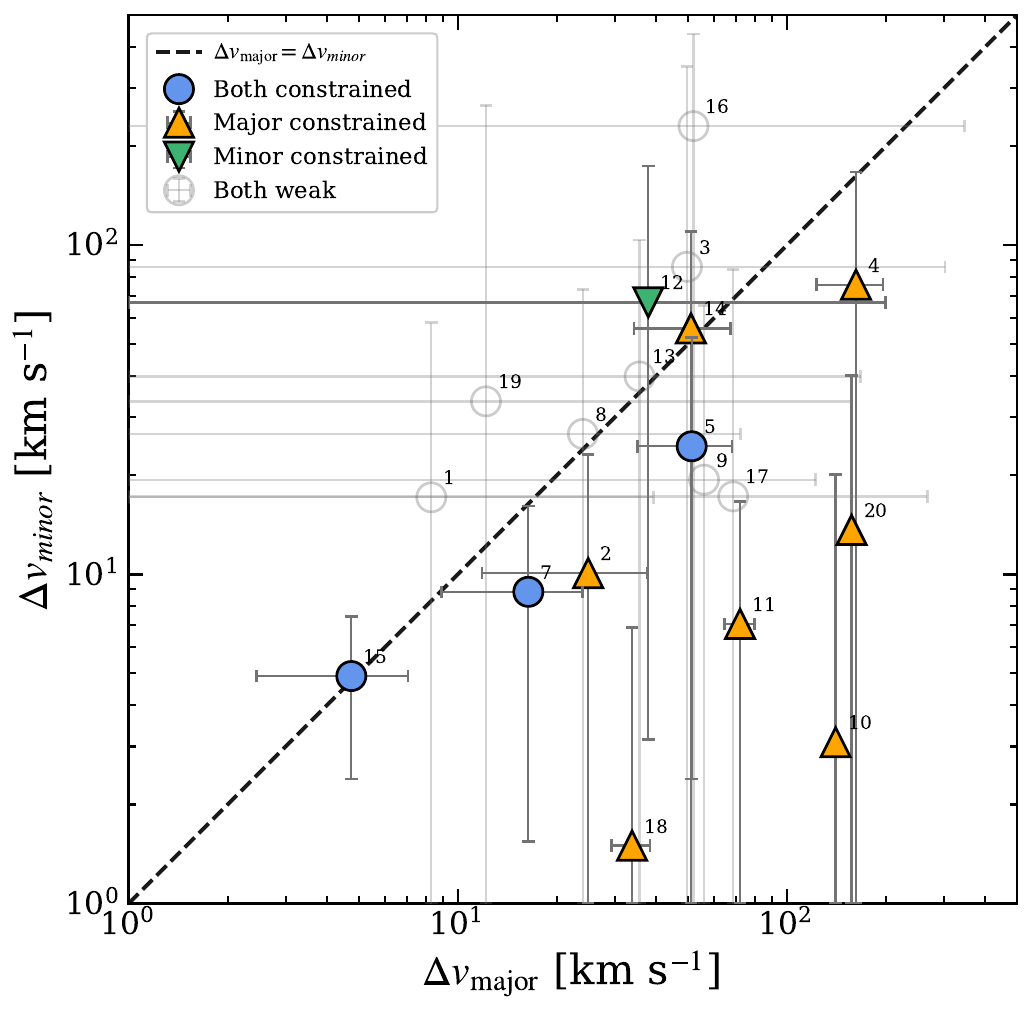}} \\
    \subfigure{\includegraphics[width = 0.49\textwidth]{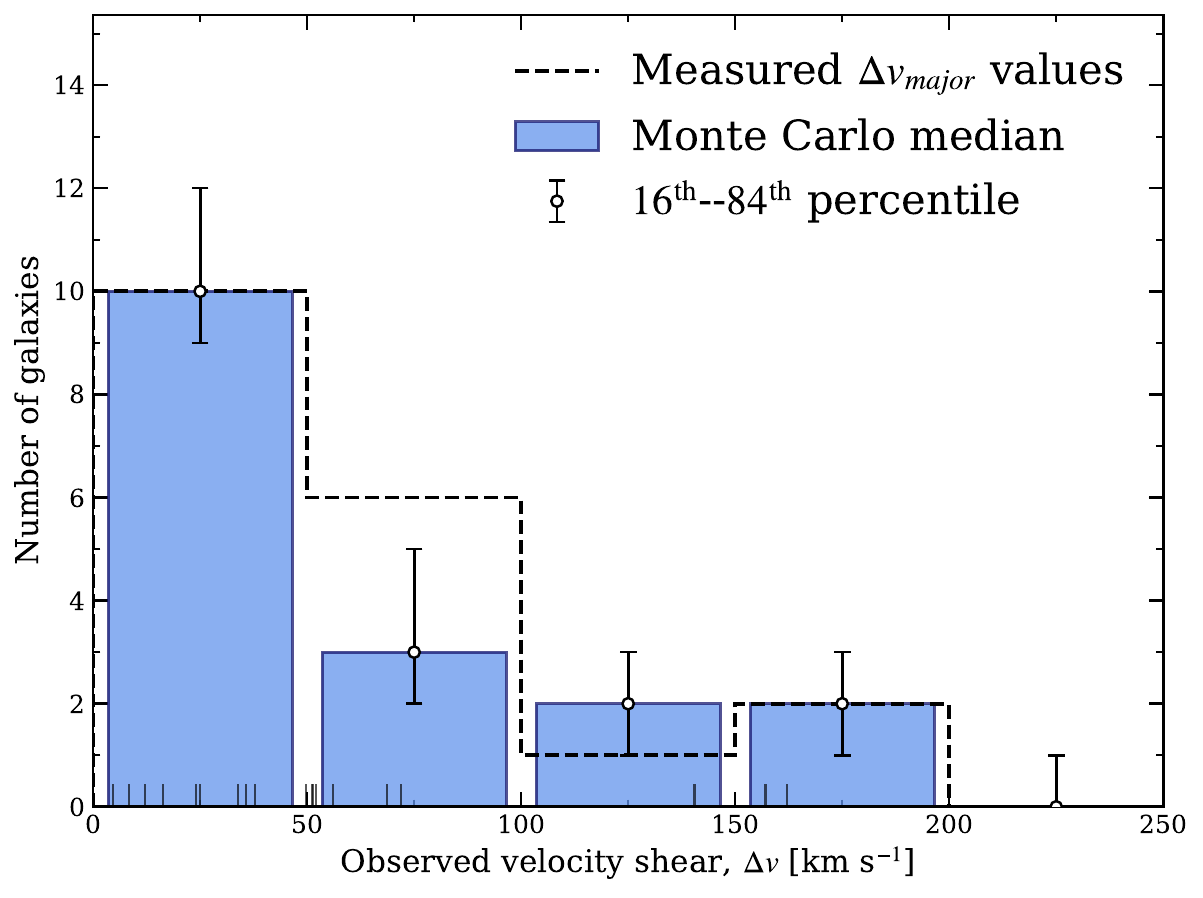}}
    \subfigure{\includegraphics[width = 0.49\textwidth]{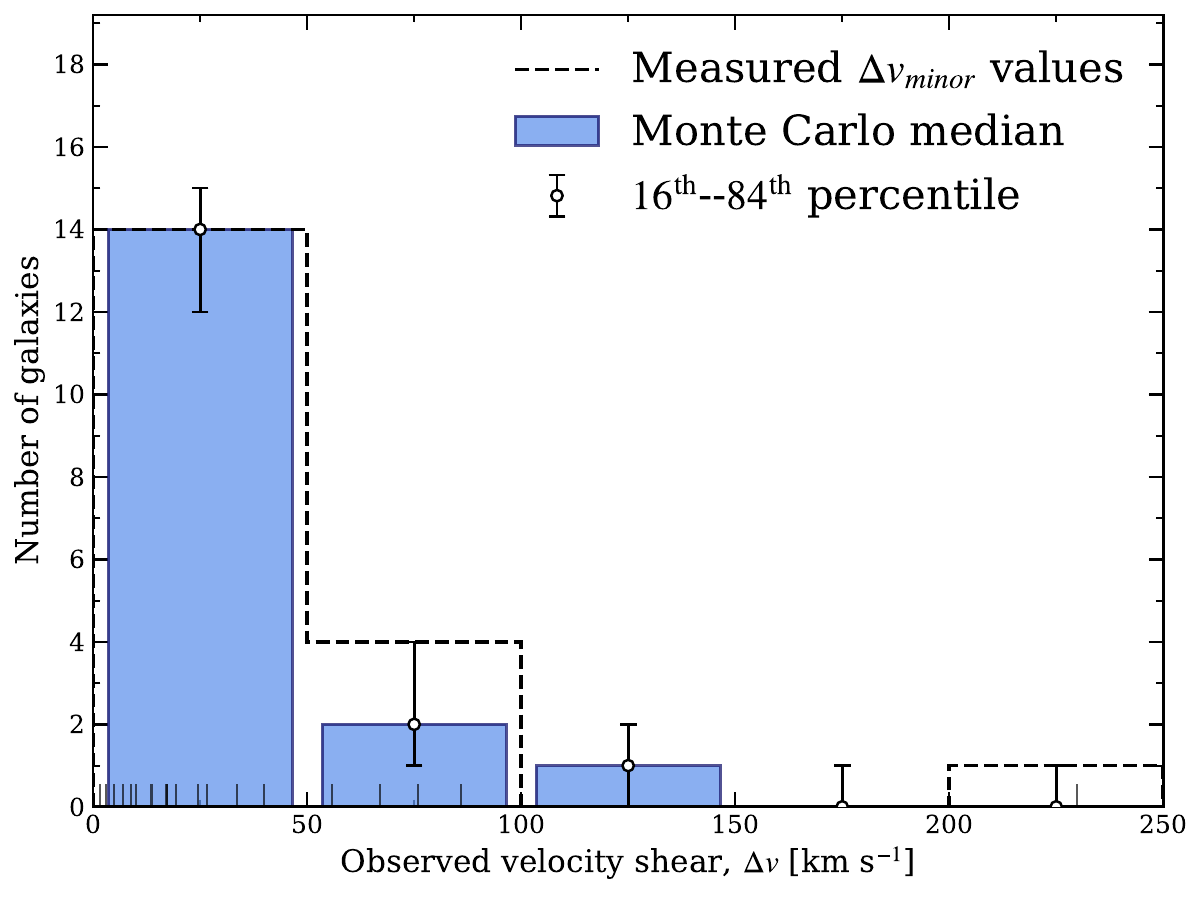}}
    \hfill
    \caption{\textbf{Top: }Comparison of the observed velocity shear measured along the morphological major axis, ($\Delta v_{\rm major}$), and along the perpendicular direction, ($\Delta v_{minor}$). The dashed diagonal line represents ($\Delta v_{\rm major}=\Delta v_{minor}$). Sources below the line have a larger velocity shear along the morphological major axis, while sources above the line show a stronger velocity variation along the perpendicular direction. Error bars represent the asymmetric uncertainties on the measured velocity shears. Galaxy IDs are indicated next to the corresponding measurements. Both axes are shown on logarithmic scales. \textbf{Bottom: }Distribution of the observed velocity shear measured from the pseudo-slit spectra. The left panel shows the velocity shear ($\Delta v$) measured along the morphological major axis, while the right panel shows the corresponding shear along the perpendicular direction. The black dashed histogram represents the Distribution obtained from the measured central values. Blue bars show the median bin counts obtained from the Monte Carlo realizations after propagating the measurement uncertainties, and the vertical error bars indicate the corresponding 16th--84th percentile range. The small marks along the bottom indicate the locations of the individual measured values ($\Delta v$).}
    \label{fig:delv_hist}
\end{figure*}

\begin{figure*}[!ht]
    \centering
    \includegraphics[width = \textwidth]{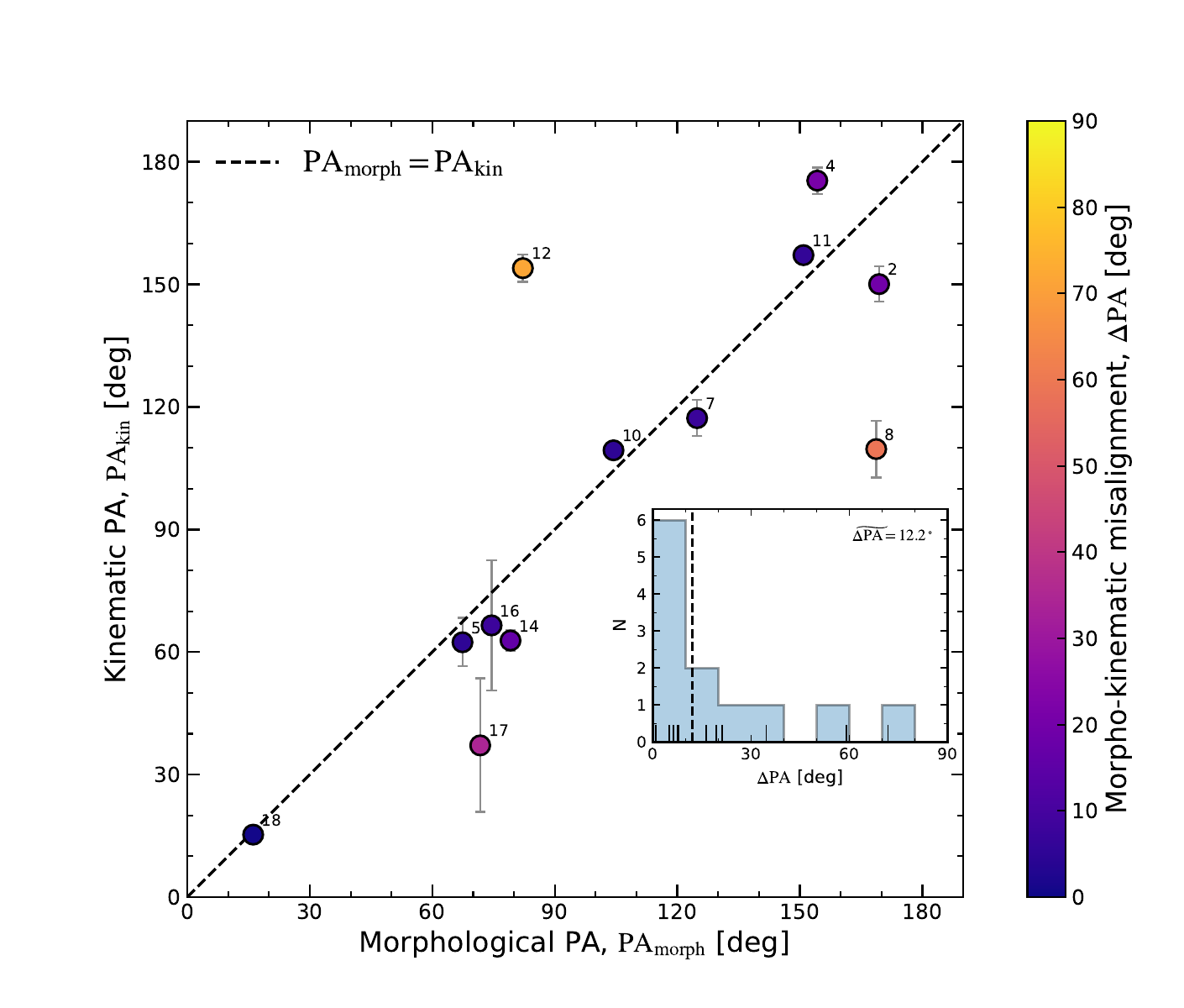}
    \caption{Comparison between the morphological position angle, $PA_{\rm morph}$, measured from the HST F775W images and the kinematic position angle, $PA_{\rm kin}$, derived from the direction of the maximum projected [O\,{\sc ii}] velocity gradient for galaxies with constrained kinematic PAs. The dashed line marks $PA_{\rm morph} = PA_{\rm kin}$. Points are colored according to the morpho-kinematic misalignment, while the error bars show the uncertainties on $PA_{\rm kin}$. Galaxy IDs are indicated next to the corresponding measurements. The inset shows the distribution of $\Delta PA$, with the dashed vertical line marking the median value of $\widetilde{\Delta PA}=12.2^\circ$. Because position angles are axial quantities, $0^\circ$ and $180^\circ$ represent the same orientation.}
    \label{fig:kin_morph_pa}
\end{figure*}

To characterize the observed spatial velocity variation without
requiring a specific dynamical model, we measure the edge-to-edge
velocity shear from the pseudo-slit profiles. For each orientation, we
define

\begin{equation}
\Delta v = \left|v_{\rm outer,1}-v_{\rm outer,2}\right|
\end{equation} \label{eqn:delta_v}

where $v_{\rm outer,1}$ and $v_{\rm outer,2}$ are the velocity
measurements at the largest reliable projected distances on opposite sides of the [O\,{\sc ii}] center. We calculate this quantity
independently along the morphological major axis,
$\Delta v_{\rm major}$, and along the perpendicular direction,
$\Delta v_{minor}$. The asymmetric uncertainties are propagated from
the corresponding uncertainties on the two outer velocity
measurements using the \texttt{add\_asym} code \citep{Barlow2003, Laursen_etal2019}.

Because several galaxies have uncertainties comparable to or larger than the measured velocity shear, we distinguish between constrained
and weak measurements. For descriptive purposes, we label a shear as
constrained when its lower propagated $68\%$ uncertainty remains above
zero, i.e. $\Delta v-\sigma_- > 0$. This classification is used only to indicate the quality of the measurement and should not be interpreted as a formal detection significance.

\begin{figure*}
    \includegraphics[width = \textwidth]{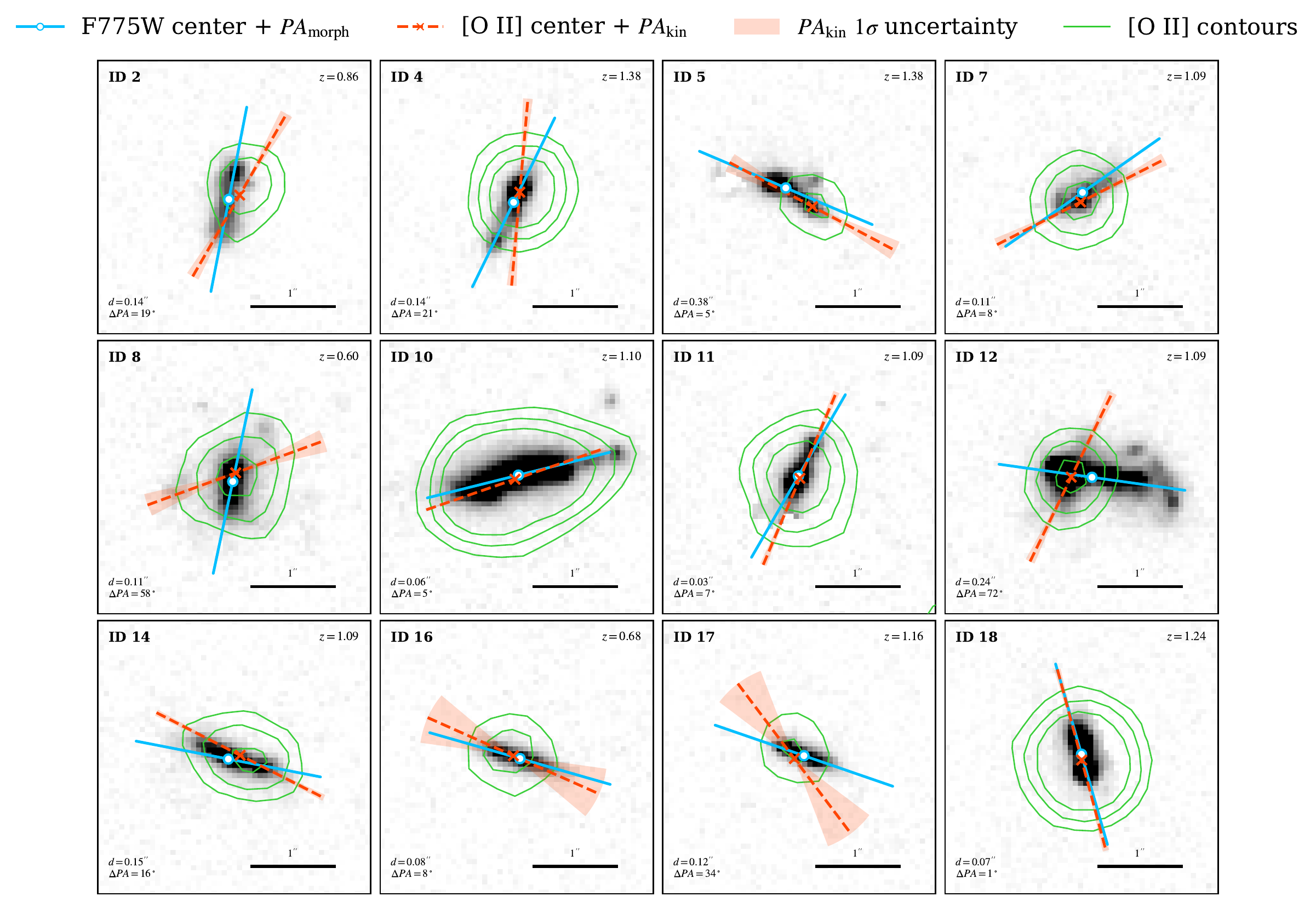}
    \caption{Morphological and ionized-gas structure of the tadpole galaxies with constrained kinematic position angles. The HST F775W cleaned images are shown, while the green contours trace the continuum-subtracted [O\,{\sc ii}] narrow-band emission. The blue open circle marks the F775W centroid and the solid blue line shows the corresponding morphological position angle, $PA_{\rm morph}$. The orange cross marks the independently measured [O\,{\sc ii}] centroid, and the dashed orange line shows the kinematic position angle, $PA_{\rm kin}$, derived from the direction of the maximum projected [O\,{\sc ii}] velocity gradient. The orange shaded sector indicates the $1\sigma$ uncertainty on $PA_{\rm kin}$. The dotted line connects the F775W and [O\,{\sc ii}] centroids, illustrating their spatial displacement. Each panel lists the galaxy ID, spectroscopic redshift, centroid separation $d$, and morpho-kinematic misalignment $\Delta PA$.}
    \label{fig:tpg_pa_kin_vs_pa_morph}
\end{figure*}

Figure~\ref{fig:delv_hist} shows the distributions of ($\Delta v_{\rm major}$) and ($\Delta v_{minor}$). The dashed histograms show the distributions obtained directly from the measured central values. To account for asymmetric measurement uncertainties, we generate Monte Carlo realizations of the velocity shear for each galaxy and reconstruct the histogram using the same fixed velocity bins across all realizations. The blue bars represent the median number of galaxies in each bin, while the vertical error bars show the corresponding 16th--84th percentile range. In both directions, most galaxies occupy the lowest velocity-shear bin, with ($\Delta v\lesssim50~{\rm kms^{-1}}$), while a smaller number of systems extend to considerably larger velocity differences. This indicates that a strong spatially resolved velocity variation is not detected in a substantial fraction of the sample, whereas a few galaxies show pronounced kinematic gradients.

Comparison between the two orientations is shown in Figure~\ref{fig:delv_hist}, where ($\Delta v_{\rm major}$) is plotted against ($\Delta v_{minor}$) for each galaxy. The dashed line marks the $1:1$ relation, ($\Delta v_{\rm major}=\Delta v_{minor}$). Galaxies lying below this line have a larger observed shear along the morphological major axis, while those above the line show a stronger velocity variation along the perpendicular direction.

A substantial fraction of the sample has velocity-shear uncertainties comparable to the measured central values, demonstrating that weak observed shear cannot be distinguished from unresolved velocity structure in many galaxies. Nevertheless, a subset of systems shows well-constrained velocity differences, including several galaxies for which the major-axis shear is more strongly constrained than the minor-axis measurement.

\section{Determination of Kinematic PA} \label{sec:kin_pa}

To determine the orientation of the dominant ionized-gas velocity gradient, we repeat the pseudo-slit extraction over a range of position angles from $0^\circ$ to $175^\circ$ in steps of $5^\circ$. For each orientation, the measured line-of-sight velocities are fitted with a linear relation,

\begin{equation}
v(r) = v_0 + g(\theta)\,r
\end{equation} \label{eqn:grad}

where $r$ is the projected distance along the slit and $g(\theta)$ is the signed velocity gradient at slit angle $\theta$.

The angular variation of the measured gradient is then fitted with
\begin{equation}
g(\theta) = a\cos\theta + b\sin\theta
          = G_{\rm max}\cos(\theta-\theta_{\rm kin}),
\end{equation}
where
\begin{equation}
G_{\rm max} = \sqrt{a^2+b^2}
\end{equation}
is the maximum projected velocity-gradient amplitude and

\begin{equation}
\theta_{\rm kin} =
\operatorname{atan2}(b,a) \bmod 180^\circ 
\end{equation}

gives the corresponding kinematic-gradient direction. The resulting angle is defined modulo $180^\circ$, since position angles are axial quantities.

The measured angle is initially obtained in the MUSE image reference frame and is converted to the sky position angle using the rotation of the MUSE WCS,
\begin{equation}
PA_{\rm kin} =
(\theta_{\rm kin}-\theta_{\rm MUSE}) \bmod 180^\circ .
\end{equation}
Uncertainties on $PA_{\rm kin}$ are estimated by Monte Carlo sampling of the measured velocity-gradient uncertainties and repeating the fit. For galaxies in which the velocity gradient is too weak to provide a well-constrained direction, we do not assign a reliable kinematic position angle.

The resulting $PA_{\rm kin}$ values are compared with the morphological position angles measured from the HST F775W images to quantify the degree of morpho-kinematic alignment.

\section{Kinematic position angles and morpho-kinematic alignment}
\label{sec:pa_alignment}

Figure~\ref{fig:kin_morph_pa} compares the kinematic position angle,
$PA_{\rm kin}$, derived from the direction of the maximum projected
[O\,{\sc ii}] velocity gradient, with the morphological position angle,
$PA_{\rm morph}$, measured from the HST F775W images. Only galaxies for
which the kinematic position angle is sufficiently constrained are
included in this comparison. The dashed line indicates
$PA_{\rm kin}=PA_{\rm morph}$.

To quantify the relative orientation of the two axes, we define the
morpho-kinematic misalignment as

\begin{equation}
\Delta PA = \min\left(|PA_{\rm morph}-PA_{\rm kin}|, 180^\circ-|PA_{\rm morph}-PA_{\rm kin}|\right)
\end{equation} \label{eqn:delta_pa}

such that $0^\circ \leq \Delta PA \leq 90^\circ$. The uncertainties on
$\Delta PA$ is obtained by propagating the uncertainties on
$PA_{\rm kin}$, while accounting for the axial nature of position angles. Such formulation is common in such studies \citep{Barrera-Ballesteros_etal2015, Ene_etal2018}. 

The distribution of $\Delta PA$ is shown in the inset of
Figure~\ref{fig:kin_morph_pa}. The galaxies with constrained
kinematic position angles have a median morpho-kinematic misalignment
of $\widetilde{\Delta PA}=12.2^\circ$. Six of the twelve systems have
$\Delta PA<10^\circ$, while nine have $\Delta PA<30^\circ$.
Thus, despite their strongly asymmetric head--tail morphologies, the
direction of the maximum projected [O\,{\sc ii}] velocity gradient is
generally aligned with the F775W morphological elongation.

A smaller subset shows substantially larger misalignments. In
particular, IDs~8, 12, and 17 have $\Delta PA\gtrsim30^\circ$,
demonstrating that the gaseous kinematic and morphological axes can
be strongly decoupled in individual tadpole systems.

\section{Morpho-Kinematic misalignment} \label{sec:morph_kin_misalignment}

\begin{figure}[!ht]
    \centering
    \includegraphics[width = 0.5\textwidth]{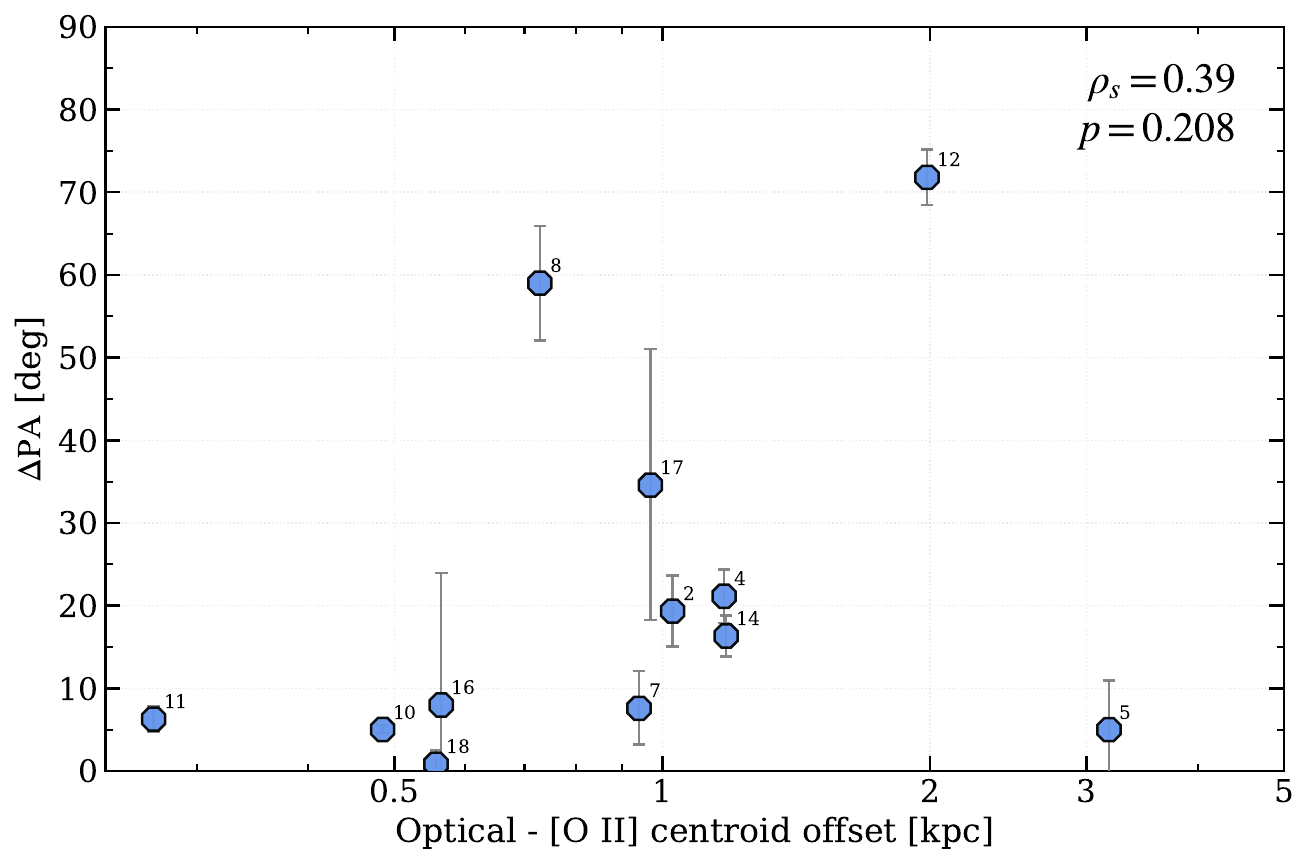}
    \caption{Morpho-kinematic misalignment, $\Delta PA$, as a function of the projected physical separation between the HST F775W and [O\,{\sc ii}] centroids for galaxies with constrained kinematic position angles. Error bars show the asymmetric uncertainties on $\Delta PA$, and galaxy IDs are indicated next to the corresponding measurements. The two quantities show substantial scatter, with a Spearman rank coefficient of $\rho_{\rm s}=0.39$ and $p=0.208$.}
    \label{fig:delta_pa_offset}
\end{figure}

We next examine whether the projected spatial displacement between the F775W and [O\,{\sc ii}] centroids are related to the degree of morpho-kinematic misalignment. Figure~\ref{fig:delta_pa_offset} shows $\Delta PA$ as a function of the projected physical centroid offset for galaxies with constrained kinematic position angles. The majority of systems with offsets below $\sim1$ kpc have relatively small
misalignments, typically $\Delta PA \lesssim 10^\circ$, although a few Objects show larger angular offsets at similar projected separations.

At larger centroid displacements, the sample extends toward larger
values of $\Delta PA$. In particular, the most strongly misaligned
system reaches $\Delta PA \sim 72^\circ$ at a projected offset of a few kiloparsecs. The relation is not one-to-one, however, and some galaxies with large spatial offsets retain only moderate morpho-kinematic misalignments.

We next examine whether the projected spatial displacement between the
F775W and [O\,{\sc ii}] centroids is related to the degree of
morpho-kinematic misalignment. Figure~\ref{fig:delta_pa_offset} shows
$\Delta PA$ (see eqn~\ref{eqn:delta_pa}) as a function of the projected physical centroid separation for the galaxies with constrained kinematic position angles.

The two quantities show considerable scatter. Galaxies with small centroid separations span a broad range of morpho-kinematic misalignments, while some systems with relatively large physical offsets retain closely aligned morphological and kinematic axes. Using the central measurements, we obtain a Spearman rank coefficient
of $\rho_{\rm s}=0.39$ with a corresponding $p$-value of $0.208$. We therefore do not find statistically significant evidence for a monotonic relation between the F775W--[O\,{\sc ii}] centroid displacement and $\Delta PA$ in the present sample.

A leave-one-out jackknife analysis was performed to assess the sensitivity of the correlation to individual galaxies. The resulting Spearman coefficients remain positive in all realizations, spanning $\rho_{\rm s}=0.25$--$0.73$, with a median value of $\widetilde{\rho_{\rm s}}\simeq0.38$. Although this indicates that the positive trend is not produced by a single object, the broad range of coefficients demonstrates that its strength is sensitive to the small
sample size. We therefore do not consider the positive correlation as statistically significant.

This suggests that spatial displacement of the ionized gas and angular misalignment of the gaseous velocity field need not occur simultaneously. The two quantities may therefore trace different aspects of the dynamical state of individual tadpole galaxies.

\section{Discussion and conclusions} 
\label{sec:conclusion}

We have studied the ionized-gas kinematics of 20 morphologically selected tadpole galaxies in the HUDF using deep VLT/MUSE spectroscopy together with HST and JWST imaging. The galaxies have a very asymmetric, elongated morphology, while their ionized gas still exhibits relatively ordered, galaxy-scale kinematics in some cases. It is worth reiterating here that a large-scale velocity gradient is consistent with rotation, but it is not uniquely produced by rotation \citep{Shapiro_etal2008, Hung_etal2015}. Interactions, mergers, gas inflows, outflows, and orbital motions of clumps can also generate coherent velocity gradients \citep{Genzel_etal2011, Rodrigues_etal2017, Simons_etal2019, Forster&Wuyts2020, deGraaff_etal2024}. This is particularly important at intermediate/high redshift, where spatial resolution and PSF effects make it difficult to distinguish a rotating disk from some merger configurations \citep{Puech2010, Davis_etal2011, Bellocchi_etal2016}. Such distinctions are relevant to the galaxies being studied here and to answering questions about how settled these galaxies really are and, more fundamentally, the actual dynamical structure of these tadpoles.

Our analyses indicate that the large-scale galaxy structure, as traced by morphology and the directions of the ionized-gas velocity field, is generally coherent, with a small offset between the morphology and the kinematic position angle ($\Delta {\text{PA}} \sim 12\ \text{degree}$). Such an alignment between photometric and kinematic axes is commonly associated with a comparatively regular large-scale velocity field. For nearby non-interacting CALIFA galaxies \citep{Barrera-Ballesteros_etal2014} found that the photometric and ionized-gas kinematic axes are generally closely aligned, even in the presence of bars. Similar photometric--kinematic alignment is also common among dynamically regular fast-rotating systems in nearby integral-field surveys \citep{Krajnovic_etal2011}. This might indicate that the asymmetric component is embedded within a larger gravitationally organized structure. To be more specific, the head-tail morphology might be a perturbation of an underlying disk-like system rather than pointing towards evidence that the entire system lacks a coherent angular momentum coherence. This is similar to what is observed among local tadpoles, where these galaxies are found to rotate despite their strongly asymmetric appearance. One must exercise caution while drawing such an inference because $PA$ alignment (rather small misalignment) itself is not a proof of dynamical equilibrium; one has to worry about projection effects (tricky in high-redshift systems with limited resolving power \citep{Pilyugin_etal2020}).

While discussing about dynamical structure and equilibrium configuration, it is worth mentioning that we are utilizing the kinematic information derived from [O\,{\sc ii}] emission which traces ionized gas from actively star-forming regions. In other words, we are measuring the kinematics of bright ionized components/knots rather than measuring the underlying stellar disk motion or cold molecular or atomic gas reservoir \citep{Falcon-Barroso_etal2006, Girard_etal2019, Girard_etal2021}. Obviously, this is relevant for tadpoles as the head is beaming with enhanced star-formation and we might obtain a biased kinematic picture of the whole galaxy. One must first answer whether the [O\,{\sc ii}] velocity field represent the kinematics of the entire galaxy or primarily of the star-forming regions \citep{Baker_etal2025}.

It appears that to unfold the actual dynamical nature of tadpoles, we must have deeper observations probing an underlying faint disk. The current observations are not adequate to do that. Because if tadpoles are simply disk galaxies viewed edge-on, we should be seeing ordered galaxy-scale motion but nearly half of the tadpoles (in our sample) show no such ordered motion or measurable (preferably strong) velocity gradient. 

To summarize, the coexistence of ordered gaseous kinematics with strongly asymmetric tadpole morphologies raises an important question regarding the relationship between morphological and dynamical disturbance in galaxies. The detection of moderate level velocity gradients in a substantial fraction of our sample suggests that the characteristic head-tail morphology does not necessarily imply a globally unsettled dynamical state. Instead, the asymmetric morphology may represent a localized or transient enhancement of star formation within an underlying galaxy-scale kinematically coherent structure.  This interpretation is supported, to some extent, by the small projected offsets between the [O\,{\sc ii}] and rest-frame optical centroids in most of our galaxies, as well as by the generally small morpho-kinematic PA misalignments. However, this might not reveal the complete picture as [O\,{\sc ii}] traces ionized gas from the star-forming regions. On the other hand, a coherent velocity gradient alone can not establish uniquely the rotation support in these galaxies. The absence of measurable ordered kinematics in the remaining systems ($\sim 35\%$) may likewise reflect either intrinsically disturbed dynamics or observational limitations, PSF effects, inclination etc. \citep{Feng_etal2022, deGraaff_etal2024}. Taken together, our analyses suggest that morphological and kinematic settlement does not occur simultaneously and these tadpole galaxies might represent a transient evolutionary phase towards the combined morpho-kinematic settlement process.    

Our primary conclusions are:\\

\begin{itemize} 

\item For a substantial fraction of the sample, the observed [O\,{\sc ii}] velocity shear is weak or poorly constrained, while a smaller subset shows clear spatial velocity gradients. For the 12 galaxies with constrained kinematic position angles, the direction of the maximum
projected [O\,{\sc ii}] velocity gradient is generally aligned with the F775W morphological elongation. The median morpho-kinematic misalignment is $\widetilde{\Delta PA}=12.2^\circ$, with nine of the
twelve systems having $\Delta PA<30^\circ$. This shows that a strongly asymmetric tadpole morphology can coexist with an organized galaxy-scale gaseous kinematic axis.

\item The F775W and [O\,{\sc ii}] centroids are also spatially close in most galaxies, with 16 of 19 systems having projected offsets below 2 kpc. However, we do not find a statistically significant correlation between centroid displacement and morpho-kinematic misalignment
($\rho_{\rm s}=0.39$, $p=0.208$), suggesting that spatial displacement and angular kinematic decoupling may trace different aspects of the galaxy dynamics.

\item In 13 out of 20 tadpole galaxies (i.e., $65\%$ in our sample), the 2D velocity maps display signatures of apparent rotation. Rest of the galaxies show no signs of rotation or ordered kinematics in the 2D velocity maps.

\item Overall, our results indicate that morphological asymmetry does not necessarily imply kinematic disorder. Kinematic organization may develop while the distribution of recent star formation remains highly
irregular, suggesting that morphological and kinematic settling need not proceed simultaneously during the assembly of low-mass galaxies. Higher spatial-resolution spectroscopy will be required to determine whether these systems are rotationally supported and whether they may
represent an early stage in the formation of later low-mass disk
galaxies.
\end{itemize}

\appendix

\section{aperture extracted spectra}

Figure~\ref{fig:spectra} shows the
observed-frame MUSE spectra of the 20 morphologically selected tadpole
galaxies. The expected locations of the principal nebular emission
lines are marked at the MUSE spectroscopic redshift of each source.
The [O\,{\sc ii}] $\lambda\lambda3726,3729$ doublet provides the
primary tracer used for the spatially resolved kinematic analysis.

\begin{figure*}[!ht]
    \centering
    \includegraphics[width = 0.9\textwidth]{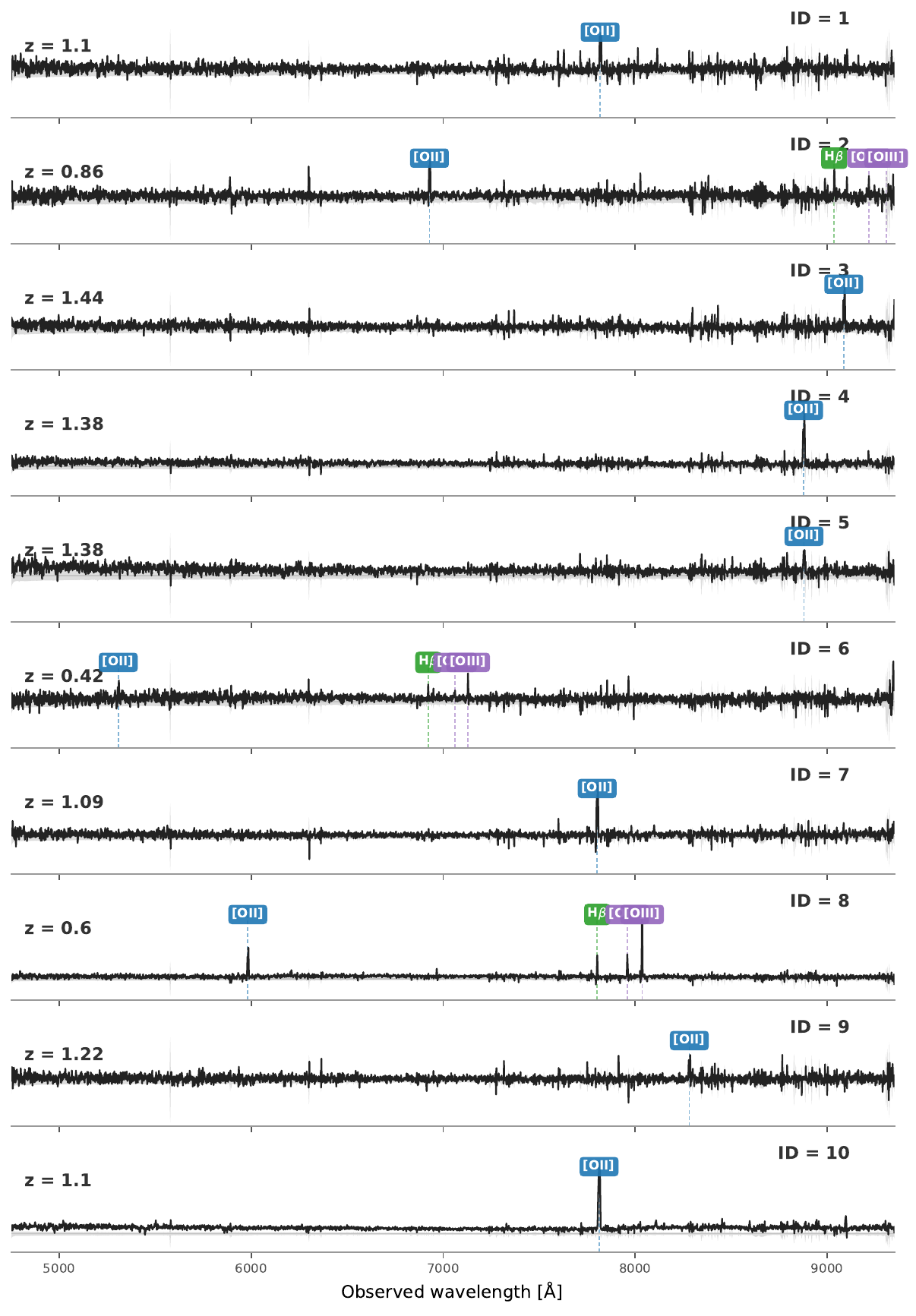} 
    \caption{Observed-frame MUSE spectra of tadpole galaxies. Each spectrum is labeled with the galaxy ID and MUSE spectroscopic redshift. The expected positions of the principal nebular emission lines are indicated by vertical dashed lines, including [O{\sc ii}] $\lambda\lambda3726,3729$, H$\beta$, and [O{\sc iii}] $\lambda\lambda4959,5007$, where covered by the MUSE wavelength range. The [O{\sc ii}] doublet is used as the primary tracer of the ionized-gas kinematics throughout this work.}
    \label{fig:spectra}
\end{figure*}

\begin{figure*}[!ht]
    \centering
    \addtocounter{figure}{-1}
    \includegraphics[width = 0.9\textwidth]{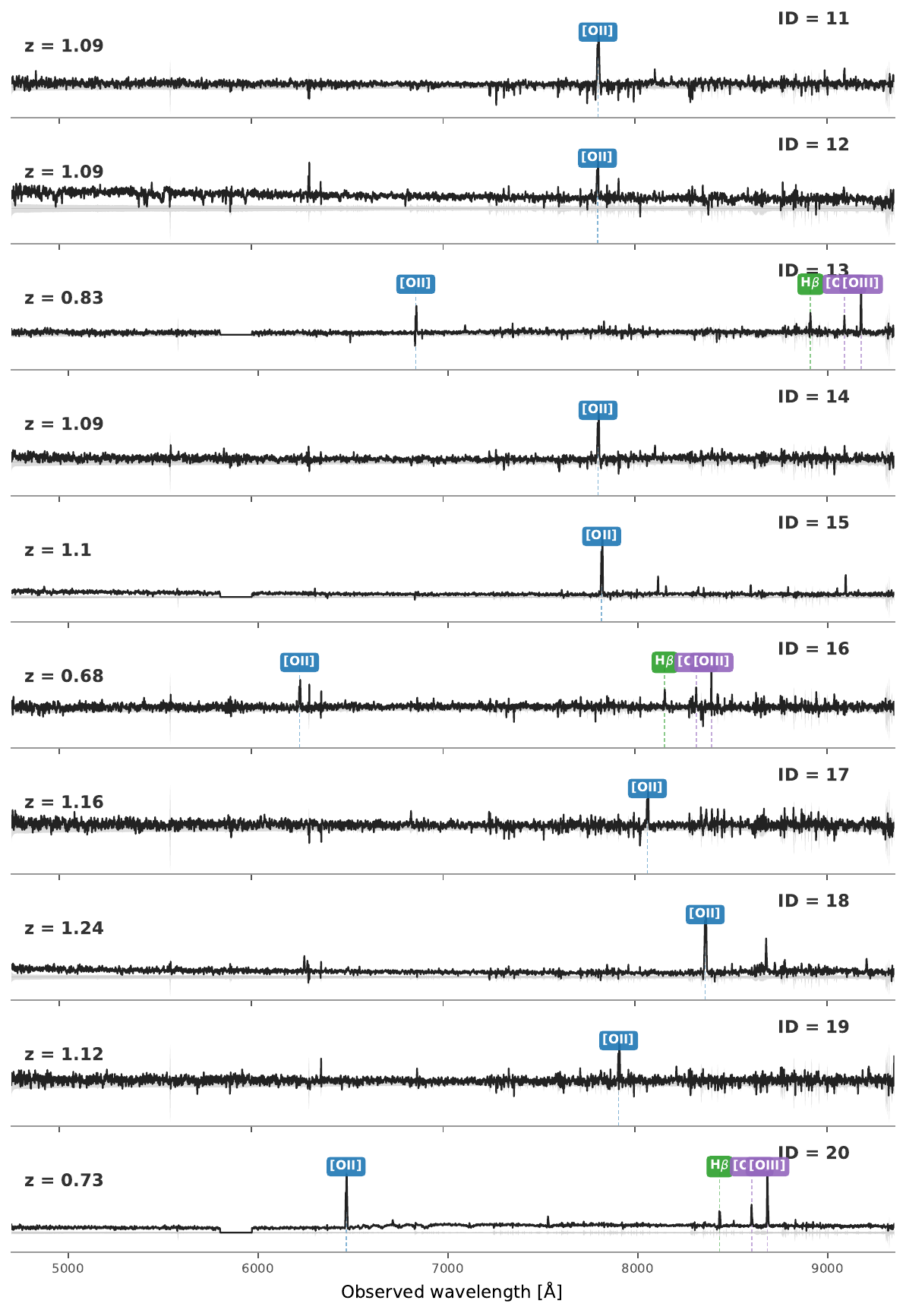}
    \caption{continuation of Figure~\ref{fig:spectra}}
\end{figure*}

\clearpage

\bibliography{refrence}{}
\bibliographystyle{aasjournal}

\end{document}